\documentclass{jfm}
\usepackage{graphicx}
\usepackage{epstopdf, epsfig}
\usepackage{graphicx}
\usepackage{newtxtext}
\usepackage{newtxmath}
\usepackage{hyperref}
\usepackage{color}
\usepackage{siunitx}
\usepackage{soul}

\newcommand{\dd}[2]{\frac{\mathrm{d} #1}{\mathrm{d} #2}}
\newcommand{\ddp}[2]{\frac{\partial #1}{\partial #2}}
\newcommand{\order}[1]{\mathcal{O}\left(#1\right)}
\newcommand{\red}[1]{}
\newcommand{\rout}[1]{}
\newcommand{\blue}[1]{#1}

\renewcommand{\Pi}{P}
\renewcommand{\Lambda}{H} %change definition of perturbed quantities

\newcommand{\poisson}{\nu} %Poisson's ratio of the channel walls
\newcommand{\aspect}{a} %aspect ratio: \aspect = [z]/[y]
\newcommand{\amplitude}{\delta} %amplitude of perturbation in scaling
\newcommand{\bendability}{\blue{\Gamma}}
\newcommand{\h}{h}
\newcommand{\x}{x}
\newcommand{\y}{y}

\newcommand\abeqn[2]{\refstepcounter{equation}
     \[
     \label{#1}
     #2
     \eqno{\text{(\theequation)}\text{a,b}}
     \]
}

\shorttitle{Bendocapillary Instability of Liquid in a Flexible-Walled Channel} 
\shortauthor{A. T. Bradley, I. J. Hewitt, D. Vella}

\title{Bendocapillary Instability of Liquid in a Flexible-Walled Channel}

\author{Alexander T. Bradley\aff{1,2}\corresp{\email{aleey@bas.ac.uk}},
  Ian J. Hewitt\aff{1},
 \and Dominic Vella\aff{1} }

\affiliation{\aff{1} Mathematical Institute, University of Oxford, Woodstock Rd, Oxford, OX2 6GG, United Kingdom
\aff{2} British Antarctic Survey, High Cross, Madingley Road, Cambridge, 
CB3 0ET, United Kingdom} 
\begin{document}

\maketitle

\begin{abstract}
We study the bendocapillary instability of a liquid droplet that part fills a flexible walled channel. Inspired by experiments in which a \red{`weaving'}\blue{periodic} pattern emerges as droplets of liquid are condensed slowly into deformable microchannels, we develop a mathematical model of this instability. We describe equilibria of the system, and use a combination of numerical methods, and asymptotic analysis in the limit of small channel wall deflections, to elucidate the key features of this instability. We find that configurations are \rout{always} unstable to perturbations of sufficiently small wavenumber \blue{regardless of parameter values}, that the growth rate of the instability is highly sensitive to the volume of liquid in the channel, and that both wetting and non-wetting configurations are susceptible to the instability in the same channel. Insight into novel interfacial instabilities opens the possibility for their control and thus exploitation in processes such as microfabrication.
\end{abstract}
%part 1: demonstrating the instability with constant volume, for wetting and non-wetting. Model + equilibria + scaling argument(?) + numerics for bvp + asymptotics for both wetting and non-wetting cases. Think we want to show experiments here as further motivation, then say here we consider only the statics (i.e. what happens if I inject liquid very slowly)

%part 2: how does a time dependent volume change the picture? Brief model + describe base states + linear stability analysis + might need to do some non-linear analysis?

\graphicspath{{./}}

\section{Introduction}\label{S:Introduction}
%Interfacial Instabilities are very common [some examples]
Instabilities at liquid-liquid interfaces play an important role in many familiar phenomena, from the break up of a water column leaving a faucet~\citep{Plateau1873, Rayleigh1879PRSL}, to the formation of trains of soap bubbles~\citep{Eggers2008RepProgPhys}, and the dendritic morphology of snowflakes~\citep{Langer1980RevModPhys}. %The traditional motivation for understanding such instabilities is to prevent the problems that they cause, such as the early failure in zinc alkaline batteries~\citep{Gallaway2010Electrochem}, the collapse of foetal airways~\citep{Halpern1992JFM}, and reduced pattern quality in inkjet printing~\citep{Calvert2007Science}. Recently, however, attention has shifted towards understanding interfacial instabilities in order to utilize them; for example, interfacial instabilities have been successfully exploited to increase the packing density of carbon nanotube forests~\citep{Chakrapani2004PNAS}, manufacture periodic assemblies~\cite[e.g.][]{DeVolder2013Angewandte}, and trap colloids at surfaces~\citep{Pokroy2009Science}.
Small scale fluid flow confined by solid boundaries is fundamental to many such situations.  On these scales, surface forces dominate over body forces, and thus capillarity plays a dominant role in controlling the flow. Capillary flows arise from the deformation of an interface, so the shape of the confinement is critical to the behaviour of such instabilities. Take, for example, the Saffman-Taylor (or `viscous fingering') instability~\citep{Saffman1958PRSL}, which classically occurs when a liquid of lower viscosity displaces a liquid of higher viscosity in a channel whose width is uniform; this instability can be suppressed by either varying the channel width in space~\citep{AlHousseiny2012NaturePhysics, AlHousseiny2013PhysFlu, Reyssat2014JFM} or in time~\citep{Zheng2015PRL}. [Indeed, channel tapering can even promote an `opposite' instability in which less viscous liquid is able to displace more viscous liquid from the apex of a wedge~\citep{Keiser2016JFM}.] On scales that are smaller still, there is evidence that evaporation driven instabilities common in drying of microelectromechanical systems have a sensitive dependence on the geometry of the channel~\citep[for example]{Hadjittofis2016JFM, LedesmaAguilar2017SoftMatter, Ha2021SoftMatter}.
%When the walls are flexible, new possibilities open up. For example, the Saffman Taylor instability can be suppressed by replacing one of the walls by a flexible walls 

\begin{figure}
    \centering
    \includegraphics[width = \textwidth]{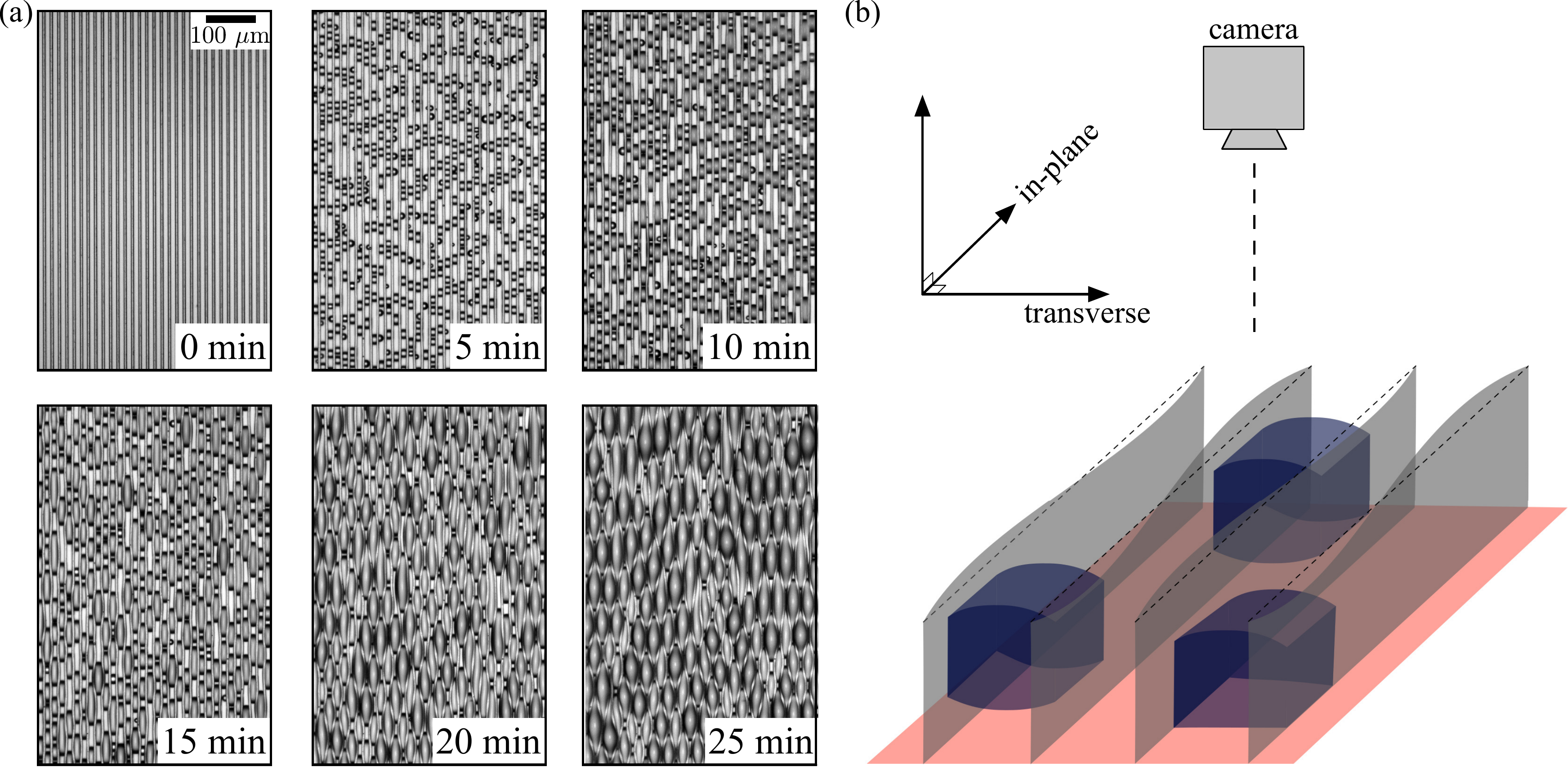}
    \caption{(a) Snapshots of experiments performed by R. Seemann (personal communication), similar to those published in ref~\cite{Seemann2011JPhysCondMat}, in which liquid droplets are condensed within an array of deformable microchannels. The interaction between the droplets and deformable boundaries leads to a pattern of drops in neighbouring channels offset relative to one another, and with a characteristic droplet spacing and size. (b) Schematic diagram of the experiments shown in (a). \blue{Here, droplets are shaded blue, flexible channel walls are shaded grey, and the base of the array is shaded red. It is not clear in the experiments whether the droplets extend to the base of the channels or not (personal communication). Black dashed lines indicate where the top of the channel walls would be located, if they were undeformed.}  We refer to variations in the \rout{plane}\blue{direction} parallel\rout{ (perpendicular)} to the channel walls as `in-plane' \blue{and those perpendicular to that direction as `transverse',} \rout{(transverse, respectively)} as indicated.}
    \label{fig:Experiments}
\end{figure}

In each of the above examples, a rigid geometry is imposed externally upon the liquid. However, new possibilities open up if the walls confining the liquid are flexible; in this case, the presence of liquid can feed back on the channel geometry, with the potential to significantly affect the behaviour of the instability. %Such fluid-structure interaction can both suppress and promote instability;  the Saffman-Taylor instability, for example, may be suppressed when one of the rigid walls is replaced by a flexible one~\citep{PihlerPuzovic2012PRL, PihlerPuzovic2013JFM}, while the instability-driven clustering of submerged micropillars that occurs when liquid is evaporated relies on the deformation of the pillars~\citep{DeVolder2013Angewandte}. 
A hint of this can be found in the experiments of~\citet{Seemann2011JPhysCondMat}, snapshots of which are shown in figure~\ref{fig:Experiments}a. In these experiments, \blue{a humid environment results in }droplets condens\rout{e}\blue{ing} into an array of deformable microchannels \rout{from a humid atmosphere}\blue{which is described by~\citet{Seemann2011JPhysCondMat} as elastic}. As condensation proceeds, the surface of the droplets moves towards the free end of the channels (the ends closest to the camera in figure~\ref{fig:Experiments}\blue{b, which corresponds to out of the page in figure~\ref{fig:Experiments}a}), and the droplets spontaneously arrange into a periodic \red{`weaving'} pattern. It is natural to suspect that the \red{weaving}\blue{periodic} pattern is the result of the growth of an interfacial instability that relies on fluid-induced deformations of the channel walls. In more detail: in these experiments, droplets of liquid condense into a three-dimensional array of channels (shown schematically in figure~\ref{fig:Experiments}b); the Laplace pressure within the droplets is positive because the channel walls are non-wetting, resulting in an outward deformation of the adjacent channel walls, which grows further as the volume of liquid continues to grow via condensation. Increases in channel wall deformation tend to promote the pattern formation (the presence of liquid only in isolated regions) by reducing the meniscus pressure \blue{at the widest points} (\rout{droplet}\blue{the} pressure is inversely proportional to channel width) thus promoting \blue{liquid flow towards those points and encouraging localisation into droplets (figure~\ref{fig:Experiments}b)}\rout{flow towards the centre of the droplet}. By contrast, if the channel walls were rigid and parallel, the meniscus pressure would be uniform and there would, therefore, be no preference for \blue{localization}\rout{flow towards the centre of the droplets}. 

\begin{figure}
    \centering
    \includegraphics[width=\textwidth]{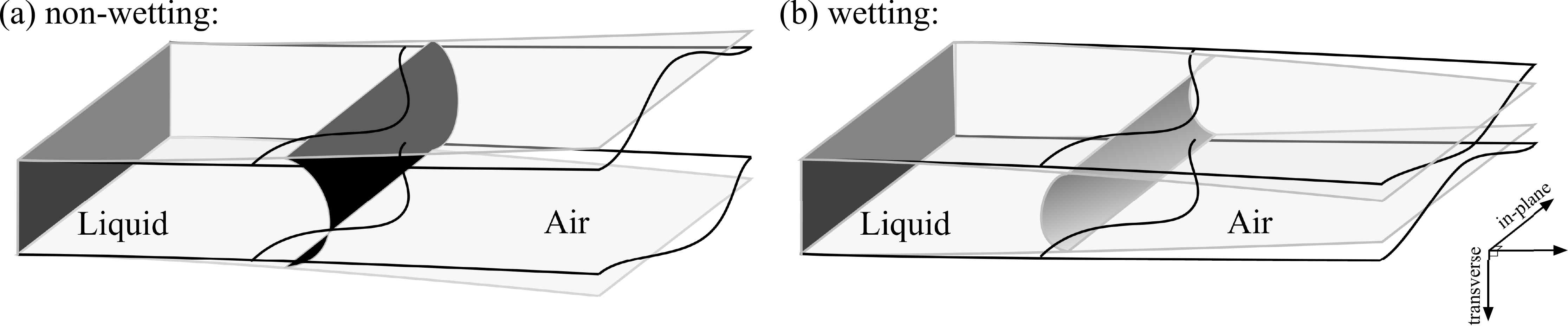}
    \caption{Schematic representation of the air-liquid interface in the bendocapillary instability for (a) non-wetting and (b) wetting liquids. In each case, the grey and black outlines indicate the configuration (channel shape and contact line) prior and post perturbation, respectively. Upon perturbing, the contact line is deformed from a straight line to a periodic curve; in both wetting and non-wetting cases, the channel experiences a deformation that enhances (reduces, respectively) the deformation of the channel walls in the base state in regions adjacent to protrusions (invaginations). \blue{Note that the configurations shown in this figure are oriented in a way that is rotated 90\si{\degree} relative to the configuration shown in figure~\ref{fig:Experiments}b.} \rout{The inset in (b) indicates the in-plane and transverse directions referred to throughout the main text.} }
    \label{fig:mechanism_schematic}
\end{figure}

The complex contact line, as well as possible interaction between droplets in neighbouring channels, adds significant complexity to a conceptual description of the\red{ weaving} instability \blue{that results in the periodic pattern} of figure~\ref{fig:Experiments}a\rout{;}\blue{.} \rout{i}\blue{I}n this paper, we consider a simplified setup, shown schematically in figure~\ref{fig:mechanism_schematic}, in which the channel is rotated\blue{ relative to figure~\ref{fig:Experiments}b} so that its `base' is on the left: a single channel consisting of two flexible walls and a rigid base, which is part filled with liquid. This setup retains the interaction between capillary flow and bending deformations (or `bendocapillarity'), and hence captures the essence of \rout{this}\blue{the} instability, whilst removing the complexity of the contact line geometry and interaction between droplets in neighbouring channels. We also consider only a constant droplet volume, simplifying the system considerably (this assumption is reasonable because condensation occurs very `slowly' in the experiments, though we can only quantify this statement in due course). \blue{We stress that we are not attempting to entirely describe the mechanism responsible for the instability shown in figure~\ref{fig:Experiments}, rather we seek to understand the fundamental aspects of instabilities resulting from channel wall deformations in two dimensions under capillary forces.}

The mechanism for instability in this setup is elucidated in figure~\ref{fig:mechanism_schematic}. The base state has a stationary interface, which is uniform in the in-plane direction\red{ [the direction parallel to the plane of the channel walls if they were undeformed, see inset in figure~\ref{fig:mechanism_schematic}(b)]}. For a non-wetting liquid with a positive Laplace pressure, the channel walls are tapered outwards (figure~\ref{fig:mechanism_schematic}a). At protrusions of a perturbation to this interface there will be a reduced confinement from the channel walls, which results from the combination of two complementary effects: (1) the interface advancing further into the channel, whose walls are tapered outwards in this direction, and (2) the elastic response of the channel walls to the advance of the interface, which takes the form of an enhanced outwards deformation (the positive Laplace pressure is now applied over a longer length). This reduction in confinement tends to reduce the liquid pressure at these protrusions (the pressure is inversely proportional to the channel width); the perturbation will grow, and the instability will be amplified, if the total pressure reduction from the combination of these two effects exceeds the (stabilizing) pressure increase that results from an interface that is now locally convex in the in-plane direction, i.e. if the wavelength of the perturbation is sufficiently long.

Perhaps surprisingly, a similar mechanism is also applicable to wetting configurations. In the wetting case, the bulk liquid pressure is negative, and the associated channel deformations are inwards, but the result is the same: protrusions now advance into a narrower channel, which is further enhanced by the additional elastic response. The combination of these two effects reduces the local liquid pressure (it becomes more negative), which can, again, cause the perturbation to grow if its wavelength is sufficiently long. Therefore, both wetting and non-wetting liquids may\rout{ theoretically} experience this bendocapillary instability in the same channel. This is in contrast to \rout{the closest analogy}\blue{the similar systems described by~\cite{AlHousseiny2012NaturePhysics} and~\citet{Keiser2016JFM} featuring rigid, tapered channel walls}\rout{in a rigid channel, in which the walls are tapered}\red{~\cite{AlHousseiny2012NaturePhysics}}. %: configurations with wetting (non-wetting, respectively) liquids are only unstable to perturbations with sufficiently long wavelengths if the channel is tapered inwards (outwards) away from the liquid.

In this paper, we investigate this instability mechanism quantitatively. We begin by presenting a simple scaling argument for the wavenumber of unstable modes and the associated growth rates, before developing a formal mathematical model of the system shown in figure~\ref{fig:mechanism_schematic}. The model equations are then non-dimensionalized; in doing so, we identify three key dimensionless parameters that characterize the aspect ratio of the channel, the ability of the liquid to deform it, and the amount of liquid it contains. Following this, in \S\ref{S:Equilibria}, we set out the base states (equilibria) of the system, which are parametrized by these three dimensionless numbers; these equilibria are essentially those described by~\citet{Taroni2012JFM}, but we extend this description to include non-wetting configurations. In \S\ref{S:LSA}, we consider the linear stability of these equilibria. We numerically solve the equations that must be satisfied by perturbations, and identify several important, yet generic features of these solutions. In particular, equilibria are linearly unstable to perturbations of sufficiently small wavenumber (large wavelength), with maximum growth rates that increase with both the channel bendability and amount of liquid in the channel. In \S\ref{S:Asymptotics}, we consider the limiting case of small channel deformations. Using asymptotic methods, we examine the system of equations that perturbations must satisfy in this limit, deriving a universal dispersion relation, and verifying analytically the observations of \S\ref{S:LSA}. Finally, in \S\ref{S:Conclusion}, we discuss and summarize the results, and provide concluding remarks.

\section{Scaling Argument for the Basic Mechanism}\label{S:Scaling}
\begin{figure}
\centering
\includegraphics[width = .99\textwidth]{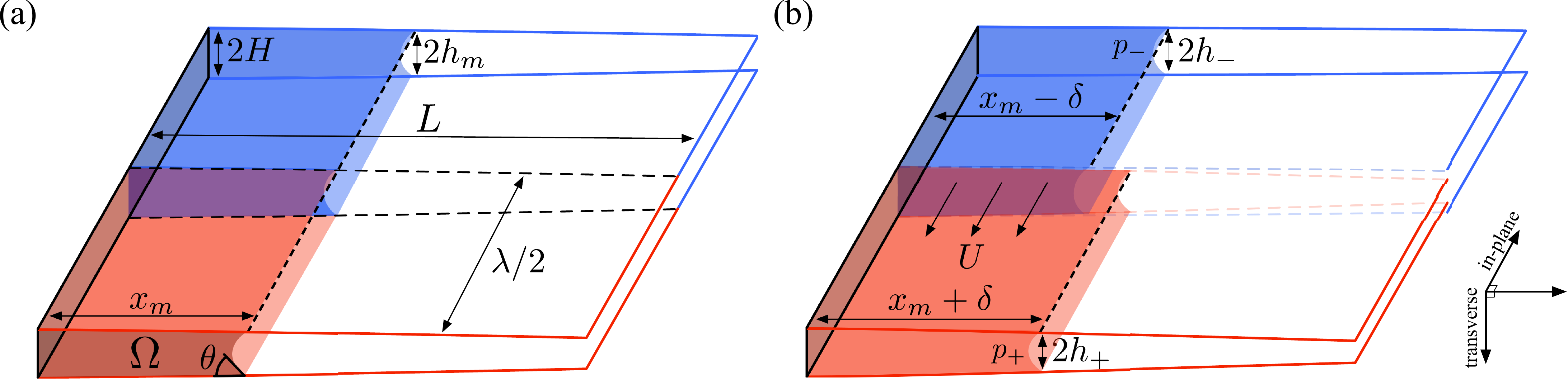}
\caption{Schematic diagrams of a section of a flexible channel consisting of a solid base and two flexible walls, which are only permitted to bend in the transverse direction. A cut along the centre of the channel (black dashed line) allows the two halves to bend independently of one another. (a) The system is in equilibrium with the meniscus located a distance $x_m$ from the base (the left-hand end, in this orientation).  (b) The equilibrium is perturbed by moving the menisci on either side of the cut a distance $\delta$; as described in the main text, if $\lambda$ is sufficiently large, this perturbation results in the flow of liquid from troughs (blue) to peaks (red) with speed $U > 0$, amplifying the perturbation. \blue{Note that the configurations shown in this figure are oriented in a direction rotated 90\si{\degree} relative to the configuration shown in figure~\ref{fig:Experiments}b.}}
\label{fig:Scaling:ScalingArgument}
\end{figure}

%gain quantitative insight into the mechanisms introduced in the introduction
Before developing a detailed mathematical model, we seek first to gain quantitative insight into the mode selection that results from the mechanism described in \S\ref{S:Introduction}. We consider the configuration shown in figure~\ref{fig:Scaling:ScalingArgument}: a section of a channel of width $\lambda$ and length $L$ contains liquid that sits adjacent to rigid base of thickness $2H$. The walls are clamped at the rigid base, while the opposite end is open. These side walls are narrow and flexible; they bend in response to the liquid pressure, and are characterized by a bending stiffness $B$. For the purposes of this scaling argument, we imagine a cut along their centre (black dashed lines in figure~\ref{fig:Scaling:ScalingArgument}a), allowing either side of this cut to deform independently of the other. We restrict deformations to the transverse direction, so that each half is itself uniform in the in-plane direction. (Note that the schematic shown in figure~\ref{fig:Scaling:ScalingArgument} corresponds to a wetting configuration, but the scaling argument set out in this section is also applicable for non-wetting configurations.)

\subsection{Base State}
\citet{Taroni2012JFM} considered the two-dimensional analogue of this system, in which both halves are identical, and showed that equilibrium configurations always exist when the cross-sectional volume of liquid $\Omega$ is sufficiently small. In particular, this means that equilibria always exist when $\Omega / (2HL) \ll 1$ and the cross-sectional volume of liquid is small in comparison with the cross-sectional channel volume. Here we consider such an equilibrium with a meniscus that is located a distance (denoted $x_m$) from the clamped end (figure~\ref{fig:Scaling:ScalingArgument}). The restriction to relatively small volumes means that the channel walls are not deformed significantly, and thus $h_m$ -- the half-width at the meniscus -- scales with the clamped end width, i.e. $h_m \sim H$. By conservation of mass, the meniscus position $x_m \sim \Omega / H$, and the liquid pressure $p_m \sim -\gamma \cos \theta/H$, where $\gamma$ is the surface tension coefficient of the liquid, and $\theta$ is the constant contact angle between the liquid and channel wall.

Before we consider perturbations to this equilibrium, we require a scaling for $h_m'$, the channel slope \rout{in the transverse direction at the meniscus}. By considering a cantilever beam with bending stiffness $B$ deformed over a length scale $x_m$ by a uniform (Laplace) pressure $-\gamma \cos \theta / H$, we find~\citep{Timoshenko1959},
\begin{equation}\label{E:Scaling:hmprimed}
h_m' \sim -\frac{\gamma \cos \theta \ x_m^3}{B H}.
\end{equation}

\subsection{Mode selection}
We mimic a periodic perturbation of wavenumber $k = 2\pi/\lambda$ and amplitude $\amplitude$, by considering a region of length $\lambda$ in the in-plane direction, centred around the cut (figure~\ref{fig:Scaling:ScalingArgument}). We imagine forcing one side of the cut to advance uniformly to $x_0 + \amplitude$ and the other to retreat to $x_0 - \amplitude$ (figure~\ref{fig:Scaling:ScalingArgument} b). As discussed, the transverse interfacial curvature, and thus liquid pressure, changes as a result of this perturbation: in this wetting example, the protruding half of the interface is forced into a stronger confinement by the tapering of the equilibrium configuration, and this confinement is enhanced by the elastic response of the channel to the change in meniscus position. \blue{Note that in this idealized scenario, the interface consists of two discrete halves, each of which is uniform in the in-plane direction, and therefore does not naturally include the stabilizing effect of in-plane curvature variations.} To account for \rout{the}\blue{this} stabilizing curvature in the context of \rout{this,}\blue{the square wave perturbation,} we add a pressure penalty of $\gamma \delta k^2$ to the protruding half \blue{(the $k^2$ term is introduced to represent the second derivative of a perturbation acting over a wavelength $1/k$, i.e.~smoothing out the square wave perturbation to be effectively sinusoidal)}.

The perturbation will grow provided that the difference in liquid pressure between the two halves, $\Delta P = p_+ - p_-$, drives liquid towards the protruding half (the red half in figure~\ref{fig:Scaling:ScalingArgument}b), i.e.~when $\Delta P<0$.

To leading order in $\amplitude$, we find that
\begin{equation}\label{E:Scaling:DeltaP_preliminary}
\Delta P\sim \gamma\left(\frac{\cos \theta}{h_-} - \frac{ \cos \theta}{h_+} + \beta \amplitude k^2\right) \sim \gamma \left(\frac{\Delta h \cos \theta}{h_m^2} + \beta \amplitude k^2\right),
\end{equation}
where $\beta>0$ is an $\mathcal{O}(1)$ scaling constant and $\Delta h = h_+ - h_- $ is the difference in the channel widths at the menisci between the two halves (see figure~\ref{fig:Scaling:ScalingArgument}b). For wetting configurations, with $\cos \theta > 0$ we expect $\Delta h < 0$, while for non-wetting configurations with $\cos \theta < 0$, we expect that $\Delta h > 0$; the first term in~\eqref{E:Scaling:DeltaP_preliminary}, which represents the transverse contribution to curvature changes, is therefore\rout{ always} negative\blue{ regardless of the wettability} and thus promotes instability.

To find a scaling for $\Delta h$, we decompose it into a contribution from the meniscus advancing into a tapered channel, and a contribution from the elastic response to the perturbation:
\begin{equation}\label{E:Scaling:ChangeInH}
\Delta h = \Delta h_\text{tapering} + \Delta h_\text{elastic}.
\end{equation}

To leading order in $\amplitude$, the tapering contribution has the same scaling as a meniscus advancing into a rigid channel whose angle is set by the equilibrium configuration, i.e.
\begin{equation}\label{E:ChangeInHtapering}
\Delta h_\text{tapering} \sim \amplitude  \times h_m' \sim -\delta \frac{\gamma  \cos \theta x_m^3}{B H},
\end{equation}
where we have used the scaling~\eqref{E:Scaling:hmprimed} for $h_m'$.

The leading order elastic contribution is found by considering a cantilever beam of length $x_m$ that is loaded with the equilibrium liquid pressure $p_m = -\gamma \cos \theta/H$ (the dry region of the beam offers no resistance to bending in this scenario~\citep{Bradley2019PRL}). If the length of this cantilever beam is then increased to $x_m + \delta$, the corresponding increase in deflection of its tip is
\begin{equation}\label{E:Scaling:ChangeInHelastic1}
\Delta h_\text{elastic} \sim -\frac{p_m}{B}\left[ \left(x_m + \amplitude\right)^4 - x_m^4\right] \red{\sim  -\amplitude \frac{\gamma \cos \theta x_m^3}{BH}}.
\end{equation}
\blue{By expanding the right hand side of~\eqref{E:Scaling:ChangeInHelastic1}, retaining only leading order terms in $\delta$, and inserting the meniscus pressure scaling $p_m \sim -\gamma \cos \theta / H$, we obtain the scaling
\begin{equation}\label{E:Scaling:ChangeInHelastic}
\Delta h_\text{elastic}\sim  -\amplitude \frac{\gamma \cos \theta x_m^3}{BH}.
\end{equation}}

Perhaps surprisingly, both the elastic~\eqref{E:Scaling:ChangeInHelastic} and tapering~\eqref{E:ChangeInHtapering} contributions to $\Delta h$\blue{~\eqref{E:Scaling:ChangeInH}} have the same scaling. Substituting these scalings into~\eqref{E:Scaling:DeltaP_preliminary}\blue{--\eqref{E:Scaling:ChangeInH}} gives
\begin{equation}\label{E:Scaling:DeltaP}
\Delta P \sim -\delta \gamma \left(\frac{\gamma \cos^2 \theta}{H^2}\frac{ x_m^3}{B H} - \beta k^2\right).
\end{equation}

By balancing the terms in~\eqref{E:Scaling:DeltaP}, we obtain
a scaling for a critical wavenumber:
\begin{equation}\label{E:Scaling:CriticalWavenumber}
k_c \sim  \left(\frac{\gamma \cos^2 \theta x_m^3}{B H^3}\right)^{1/2}
\end{equation}
We expect that perturbations with wavenumber $k \lesssim k_c$ will be unstable, while those will wavenumber $k \gtrsim k_c$ will be damped. 

\subsection{Growth rates}
When $\Delta P < 0$, liquid is sucked from the invaginations into protrusions with a typical velocity $U$ (figure~\ref{fig:Scaling:ScalingArgument}b), whose scaling we now consider. To estimate this velocity scale, we note that this pressure difference acts over a length scale $\lambda = 2\pi/k$ and so lubrication theory~\citep{Leal2007} suggests that (provided $\lambda,L \gg H$)
\begin{equation}
U \sim -\frac{H^2}{\mu}\frac{\Delta P}{\lambda}\sim \frac{H^2}{\mu}  \delta \gamma k\left(\frac{\gamma \cos^2 \theta}{H^2}\frac{x_m^3}{B H} - \beta k^2\right),
\end{equation}
where $\mu$ is the viscosity of the liquid. The corresponding flux of liquid between the two halves of the channel is
\begin{equation}\label{E:Scaling:Flux}
Q \sim \Omega U \sim\frac{H^2}{\mu}  \delta \gamma k \Omega\left(\frac{\gamma \cos^2 \theta}{H^2}\frac{ x_m^3}{B H} - \beta k^2\right),
\end{equation}
while conservation of mass for either section requires
\begin{equation}\label{E:Scaling:MassCons}
H \lambda \dd{\delta}{t} \sim Q.
\end{equation}
Combining~\eqref{E:Scaling:Flux} and~\eqref{E:Scaling:MassCons} gives a scaling for $\sigma$, the growth rate of perturbations, as
\begin{equation}\label{E:Scaling:amplitude_ode}
\sigma = \frac{1}{\delta} \dd{\delta}{t} \sim \frac{\Omega H \blue{\gamma}}{\mu}\left(\frac{\gamma \cos^2 \theta}{H^2}\frac{x_m^3}{B H} - \beta k^2\right)k^2.
\end{equation}

The wavenumber of the fastest growing modes is determined by a balance of the bracketed terms in~\eqref{E:Scaling:amplitude_ode} and can be seen to yield $k \sim k_c$, with $k_c$ as given in~\eqref{E:Scaling:CriticalWavenumber}. The corresponding growth rate of perturbations with wavenumbers of this characteristic size \rout{is}\blue{scales as}
\begin{equation}\label{E:Scaling:SigmaScaling1}
\sigma_c =  \frac{1}{\delta} \dd{\delta}{t} \sim \red{\frac{\Omega H \blue{\gamma}}{\mu}\left(\frac{\gamma \cos^2 \theta}{H^2}\frac{x_m^3}{B H} - \beta k_c^2\right)k_c^2 \sim} \frac{\gamma^{\red{2}\blue{3}} \cos^{\red{2}\blue{4}} \theta ~ x_m^7}{\mu B^{\blue{2}} H^{4}},
\end{equation}
where \rout{the final scaling uses}\blue{we have made use of} the small deformation volume scaling, $\Omega \sim x_m H$.

\blue{The dispersion relation~\eqref{E:Scaling:amplitude_ode} can be expressed as $\sigma\sim (k_*^2-k^2)k^2$ for some $k_*$. This form of the dispersion relation as a function of $k$ is reminiscent of that for the Rayleigh-Plateau instability, which describes the surface-tension driven breakup of a liquid jet falling under its own weight~\citep{Rayleigh1879PRSL, Rayleigh1892PhilosMag, Plateau1873}. However, the important distinction between the dispersion relation~\eqref{E:Scaling:amplitude_ode}, and that arising in the Rayleigh-Plateau analysis, is that the $-k^4$ in~\eqref{E:Scaling:amplitude_ode} is a result of bending stiffness, rather than surface tension, from which the $-k^4$ arises in the Rayleigh-Plateau analysis.}
 
While the above calculations are rough, they suggest that both wetting and non-wetting configurations of sufficiently small wavenumber will be amplified, and that both the fastest growing mode and corresponding growth rate are symmetric under a reversal of wettability ($\cos \theta \to -\cos \theta$). Moreover, the growth rate of unstable modes has an extremely sensitive dependence on the amount of liquid in the channel (via the meniscus position $x_m$). We now turn to a more formal calculation to interrogate these observations in detail, but shall refer back to the results of this section in due course, since this argument distils our main results.

\section{Mathematical Model}
\begin{figure}
\centering
\includegraphics[width=0.99\textwidth]{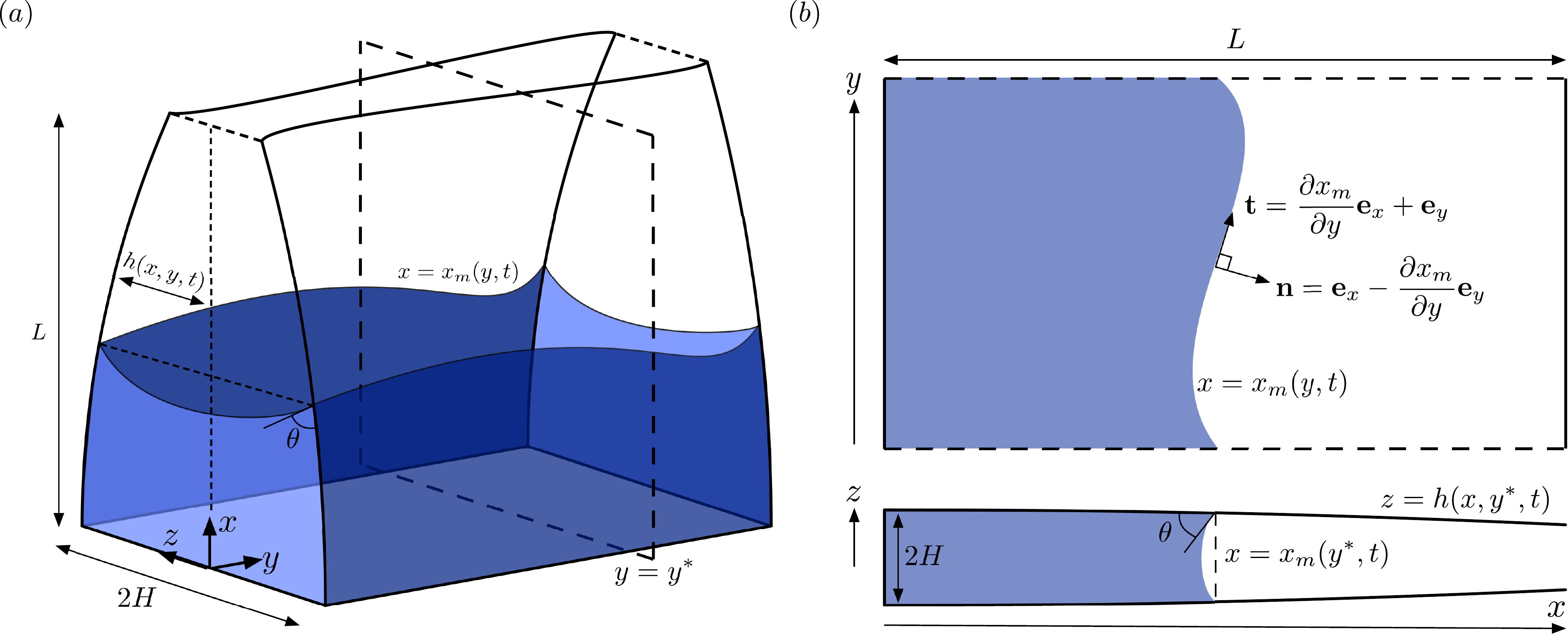}
\caption{(a) Schematic diagram of liquid in a channel consisting of a solid, impenetrable base at $x =0$ and two flexible walls, whose mid-planes are located at $z = \pm h(x,y,t)$. The liquid \blue{(a wetting liquid in this case)} makes contact with the channel walls at the contact line $x = x_m(y,t)$. The cell extends infinitely in the $y$-direction, only a section of which is shown. (The channel is assumed narrow, $H / L \ll 1$ but here we exaggerate $H/L$ for clarity.) (b) Cross sections of the system shown in (a) through the $(x,y)$ plane (upper) and $(x,z)$ plane (lower); the latter is taken through $y = y^*$, indicated by the dashed box in (a).}
\label{fig:Modelling:Schematic}
\end{figure}

%describe the model
In this section, we develop a formal mathematical model of the system discussed in \S\ref{S:Scaling}. The configuration is shown in figure~\ref{fig:Modelling:Schematic}a: a narrow cell of thickness $2H$ and length $L$ extends infinitely in the $y$-direction. The channel has a rigid boundary at $x = 0$ and is free at $x = L$. The other two walls are flexible and, when undeformed, coincide with the planes $z = \pm H$. These channel walls are characterized by their thickness $b$, Young's modulus $E$, and Poisson's ratio $\poisson$. We shall assume that channel walls are relatively thin ($b \ll L$), and may therefore be characterized by their bending stiffness $B = Eb^3 / [12(1-\poisson^2)]$. We shall take $\poisson = 0.5$, corresponding to incompressible walls, throughout.

%configuration description: liquid
Liquid of viscosity $\mu$ and surface tension $\gamma$ sits at the solid base of the channel. We assume that the liquid makes a constant contact angle $\theta$ with the channel walls (i.e. any dynamic contact angle effects are ignored). The liquid pressure induces a deformation of the channel walls; in the following sections we describe the coupled models for the flow of liquid and deformation of the channel walls, and then non-dimensionalize the resulting system of equations.

\subsection{Preliminaries and assumptions}
We assume that the configuration is symmetric about $z = 0$, and therefore only need to consider a single channel wall. Alongside our assumption that the flexible channel walls are thin in comparison with their length, this symmetry assumption means that we can characterize the channel width at time $t > 0$ entirely by the position of the mid-plane of one wall~\citep{Reddy2006}, which is denoted by $h(x,y,t)$.
%We also assume that the channel walls are thin in comparison with the channel half-width, $b \ll H$, so it is reasonable to consider $2h(x,y,t)$ to be the width of the cavity between the walls.

The channel is wetted over the region $0 < x < x_m(y,t)$. We assume that $x_m \gg H$ throughout, and also that variations in the flow in the $y$-direction occur on a length scale much longer than $H$, allowing us to use lubrication theory to model the liquid flow. Since the channel geometry does not provide a natural lengthscale for flow in the $y$-direction, we postpone discussion of the $y$-lengthscale until \S\ref{S:Modelling:NonDim}, and verify this latter assumption a posteriori. With these assumptions, the contact angle between liquid and solid is approximately that measured in the $(x,z)$ plane (figure~\ref{fig:Modelling:Schematic}b, bottom panel).

Finally, we neglect the weight and inertia of both the liquid and the channel walls, as well as the line force associated with surface tension, which have been shown to be unimportant in comparison with the Laplace pressure of the liquid in similar situations~\citep{Taroni2012JFM, Bradley2019PRL, BradleyPhDthesis}.

\subsection{Liquid flow}
The fluid flow is described using lubrication theory. The evolution of the pressure field $p(x,y,t)$ and the channel width $2h(x,y,t)$ are then coupled via Reynolds' equation
\begin{equation}\label{E:Model:Liquid:Reynolds}
\ddp{h}{t} = \nabla \cdot \left( \frac{h^3}{3\mu} \nabla p\right) \qquad \text{in}~0 < x < x_m(y,t).
\end{equation}

The free boundary of the liquid moves in response to the flux of fluid there. Ignoring any condensation or evaporation, the flux of fluid through the menisci must balance that caused by motion and thus the following kinematic condition must hold:
\begin{equation}\label{E:Model:Liquid:Kinematic}
\ddp{x_m}{t} = -\left.\frac{h^2}{3\mu}\nabla p .\frac{\mathbf{n}}{|\mathbf{n}|}\right|_{x = x_m(y,t)}
\end{equation}
Here $\mathbf{n} = \mathbf{e}_x -\partial_y x_m  \mathbf{e}_y$ is the normal to the interface in the $(x,y)$ plane (figure~\ref{fig:Modelling:Schematic}b).

According to Laplace's law, the liquid pressure adjacent to the interface is
\begin{equation}\label{E:Model:Liquid:LaplaceBC}
 p(x = x_m, y,t) = \gamma\left(C_{\perp} + C_{\parallel}\right),
\end{equation}
where $C_{\perp}$ and $C_{\parallel}$ are the transverse and in-plane interfacial curvatures, respectively. Our neglect of gravity means that the meniscus is a minimal surface: in the $(x,z)$ plane the menisci are therefore approximately arcs of circles (figure~\ref{fig:Modelling:Schematic}b) with curvatures
\begin{equation}\label{E:Model:Liquid:CPerp}
C_{\perp} = -\frac{\cos \theta}{h(x=x_m,y,t)}.
\end{equation}
Our assumption that variations in the $y$-direction occur on a length scale much larger than $H$ means that we can approximate the in-plane interfacial curvature by
\begin{equation}\label{E:Model:Liquid:CParallel}
C_{\parallel} = -\ddp{^2 x_m}{y^2}.
\end{equation}

Finally, we impose a no-flux condition at the impenetrable base:
\begin{equation}\label{E:Model:Liquid:nofluxBC}
\ddp{p}{x}=0 \qquad \text{at}~x =0.
\end{equation}

\subsection{Wall deformation}
\red{The energy penalty associated with stretching the thin channel walls, which scales with $b/L \ll 1$, is very high in comparison with the energy penalty associated with bending them, which scales with $(b/L)^3$~\citep{Pini2016SciRep}. It is therefore reasonable to neglect stretching of the channel walls, and model them simply as thin plates undergoing pure bending deformations under the applied load $p(x,y,t)$. The position of the channel wall mid-plane $h(x,y,t)$ therefore satisfies~\citep{Timoshenko1959}}
\blue{The channel walls are modelled as thin plates, and the position of their mid-planes is $z = \pm (h(x,y,t) + b/2)$, where $b$ is the wall thickness.  The bending of the walls in response to the fluid pressure satisfies}
\begin{equation}\label{E:Model:Wall:Bilaplacian}
B\nabla^4 h = p,
\end{equation}
where $p$ is the liquid pressure in $0 < x < x_m(y,t)$, and zero in $x_m(y,t) < x < L$. 

\blue{In fact, the wall deformation results from both bending and stretching of the thin plates, and can be modelled more generally by the F\"{o}ppl-von K\'{a}rm\'{a}n equations~\citep{Karman1907,Foppl1921}, which include additional second derivatives of $h$ multiplying the in-plane tension in the plates.  There are two possible sources of this tension: firstly, the `base-state' tension that arises even in the case of a uniform ($y$-independent) deflection, which is caused by the tangential component of the capillary force acting across the contact line, and, secondly the tension induced by the deformation of the wall itself.  Both of these can be considered negligible, for the following reasons.  Firstly, since the free ends of the walls are stress free, any base-state tension is introduced only over the length of the wetted portion of the walls, which scales with $L$. The size of this tension scales with the surface tension $\gamma$; the resulting terms in~\eqref{E:Model:Wall:Bilaplacian} are negligible compared to the pressure term (of order $\gamma/h$) provided $h/L \ll 1$, which we have already assumed in our application of lubrication theory.  Secondly, the base-state deflection of the walls (referred to later as $h_e(x)$) is $y$-invariant and thus involves only bending, not stretching, and therefore does not itself induce any tension in the walls.  However, $x$- and $y$-dependent perturbations to this base state do induce non-zero curvature in two dimensions, and necessarily requires stretching.  This stretching is proportional to the square of the additional deflection $\tilde{h}$, and the induced tension is of order $Eb(\tilde{h}/L)^2$.  Since we are only interested in small perturbations (indeed, we only study the linear stability problem in which the perturbations $\tilde{h}$ are formally infinitesimal) the resulting contributions to~\eqref{E:Model:Wall:Bilaplacian} will also be negligible.  We anticipate that for larger spatially-variable deflections of the walls (such as those that perhaps occur in the experiments in figure 1), the deformation-induced tension may indeed become significant, and the wall deflections would need to be modelled with the full F\"{o}ppl-von K\'{a}rm\'{a}n equations. However, for the purposes of this study we restrict ourselves to the simpler approximation~\eqref{E:Model:Wall:Bilaplacian}.}

To close the problem, we require boundary conditions at the channel ends, $x= 0$ and $x = L$, as well as at the interface, $x = x_m$. We apply a straightforward clamped condition at $x = 0$,
\begin{equation}\label{E:Model:Wall:ClampedBC}
h= H, \quad \ddp{h}{x} = 0, \qquad \text{at}~x = 0.
\end{equation}

The channel walls are free at $x = L$; for a thin plate undergoing pure bending deformations, free boundary conditions are imposed by requiring~\citep{Timoshenko1959}
\begin{equation}\label{E:Model:Wall:FreeEndBC}
\ddp{^2 h}{x^2} + \poisson \ddp{^2 h}{y^2} =0, \quad  \ddp{^3 h}{x^3} + (2-\poisson) \ddp{^3 h}{x \partial y^2} = 0\quad \text{at}~x = L.
\end{equation}

%Comment on the presence of y derivatives meaning we can't integrate out the dry region, in general?
The boundary condition~\eqref{E:Model:Wall:FreeEndBC} applies only when the channel walls do not touch. If the channel walls touch -- as might ultimately occur in the wetting case, where the walls are drawn towards one another -- the boundary condition~\eqref{E:Model:Wall:FreeEndBC} must be modified to include a repulsive shear force. This touching ends scenario requires significant deformations, which are not consistent with the relatively small volumes that are of primary interest here; we therefore assume that~\eqref{E:Model:Wall:FreeEndBC} holds throughout. 

At the meniscus, we assume that the channel and its slope, as well as the moments and shear forces it supports are continuous:
\begin{align}
\left[ h \right]_{-}^{+}  &= 0 & &\text{[continuous channel shape]},\label{E:Model:Wall:ContinuityBC1}\\
\left[\ddp{h}{x} \right]_{-}^{+} = \left[\ddp{h}{y} \right]_{-}^{+} &=0 & &\text{[continuous channel slope],}  \\ 
\left[\ddp{^2 h}{x^2} + \poisson \ddp{^2 h}{y^2} \right]_{-}^{+} =   \left[\ddp{^2 h}{y^2} + \poisson \ddp{^2 h}{x^2}  \right]_{-}^{+} =  \left[\ddp{^2 h}{x \partial y} \right]_{-}^{+} &= 0 & &\text{[continuous bending moments],} \\
\left[\ddp{^3 h}{x^3} + \ddp{^3 h}{ x\partial y^2}\right]_{-}^{+} =  \left[\ddp{^3 h}{ x^2 \partial y} + \ddp{^3 h}{y^3} \right]_{-}^{+} &=0  & &\text{[continuous shear forces].}\label{E:Model:Wall:ContinuityBC4} 
\end{align}
Here $\left[ f \right]_-^+ = f(\x_m^+,y,t) - f(\x_m^-,y,t)$ denotes the jump in the quantity $f$ across $x = x_m$, and has both $y$- and $t$-dependence in general.

In summary, the system is described by the liquid pressure $p$ and wall deflection $h$, which satisfy the coupled PDEs~\eqref{E:Model:Liquid:Reynolds} and~\eqref{E:Model:Wall:Bilaplacian} for the fluid flow and wall deformation, respectively. This system of PDEs is to be solved alongside the kinematic condition conditions~\eqref{E:Model:Liquid:Kinematic}, the no-flux condition~\eqref{E:Model:Liquid:nofluxBC}, the clamped end condition~\eqref{E:Model:Wall:ClampedBC}, the free end condition~\eqref{E:Model:Wall:FreeEndBC}, and the continuity conditions~\eqref{E:Model:Wall:ContinuityBC1}--\eqref{E:Model:Wall:ContinuityBC4}. We now turn to a non-dimensionalization of this system.

\subsection{Non-dimensionalization}\label{S:Modelling:NonDim}
The system of model equations is non-dimensionalized by scaling variables appropriately. Variations in the $x$- and $z$-directions have natural length scales set by the channel geometry, and corresponding dimensionless variables (denoted by hats)
\begin{equation}\label{E:Modelling:NonDim:SpatialScaling}
\hat{h} = \frac{h}{H}, \qquad \hat{x} = \frac{x}{L}, \qquad \hat{x}_m = \frac{x_m}{L}.
\end{equation}

The channel geometry does not provide a length scale for variations in the $y$-direction, so we choose the scale $L$ used in the $x$-direction, introducing
\begin{equation}\label{E:Modelling:NonDim:yscaling}
\hat{y} = \frac{y}{L}.
\end{equation}
Note that in the scaling argument of \S\ref{S:Scaling}, we identified a critical wavelength for instability in the $y$-direction,
\begin{equation}\label{E:Modelling:NonDim:LengthscaleFromScaling}
L_y = \left(\frac{B H^3}{\gamma  \cos^2 \theta L^3}\right)^{1/2},
\end{equation}
and so we anticipate the appearance of the dimensionless wavenumber
\begin{equation}\label{E:Modelling:NonDim:ExpectedWavenumber}
L/L_y =  \left(\frac{\gamma \cos^2 \theta L^5}{B H^3}\right)^{1/2}
\end{equation}
in our stability analysis.

Time and pressure variables are non-dimensionalized by scaling with a capillary time scale and a bending pressure scale, respectively,
\begin{equation}\label{E:Modelling:NonDim:TimeAndPressureScaling}
\hat{t} =\frac{t}{ \tau_c} =  \frac{|\gamma \cos \theta| H}{\mu L^2}t, \qquad \hat{p} = \frac{L^4}{B}p.
\end{equation}
Using values from~\citet{Bradley2019PRL}, who studied a similar bendocapillary system, the typical capillary timescale $\tau_c$ is on the order of 10~s; this is significantly shorter than the timescale on which the channels in the motivating experiments (figure~\ref{fig:Experiments}a) fill via condensation, which is on the order of minutes. The constant-volume model considered here is therefore appropriate for these experiments.

After applying the scalings above, the PDE system~\eqref{E:Model:Liquid:Reynolds} and~\eqref{E:Model:Wall:Bilaplacian} becomes
\begin{align}
\ddp{\hat{h}}{\hat{t}} &=\frac{1}{3|\bendability|}\hat{\nabla}. \left[\hat{h}^3\hat{\nabla}\hat{p}  \right] & & 0 <  \hat{x} < \hat{x}_m(\hat{y},\hat{t}),\label{E:Modelling:NonDim:PDE1}\\
  \hat{p}&=0 & &
\hat{x}_m(\hat{y},\hat{t}) < \hat{x} < 1,\label{E:Modelling:NonDim:PDE2}\\
\hat{p} &=\hat{\nabla}^4 \hat{h}& &0 < \hat{x} < 1.\label{E:Modelling:NonDim:PDE3}
\end{align}
Here
\begin{equation}\label{E:Modelling:NonDim:Bendability}
\bendability = \frac{\gamma \cos \theta L^4}{B H^2}
\end{equation}
is the channel `bendability', and characterizes the ability of the typical capillary pressure within the liquid to bend the channel walls: large $|\bendability|$ indicates that the channel walls are easily deformed (achieved by having low bending stiffness or high surface tension, for example), and vice versa for small $|\bendability|$. The sign of $\bendability$ captures the wetting conditions: $\bendability > 0$ corresponds to wetting conditions ($\cos \theta >0$), while $\bendability < 0$ corresponds to non-wetting conditions ($\cos \theta <0$).

Note that the sixth order PDE that results from combining~\eqref{E:Modelling:NonDim:PDE1}--\eqref{E:Modelling:NonDim:PDE3} is a two-spatial-dimension analogue to PDEs that often appear in studies of fluid structure interactions at low Reynolds number~\citep[see][for example]{Flitton2004EJApplMech, Duprat2011JFM, Aristoff2011IntJNonlinMech}.

After non-dimensionalizing, the boundary conditions~\eqref{E:Model:Wall:ClampedBC}--\eqref{E:Model:Wall:FreeEndBC} and~\eqref{E:Model:Wall:ContinuityBC1}--\eqref{E:Model:Wall:ContinuityBC4} on the channel wall position become
\begin{equation}\label{E:Modelling:NonDim:ClampedBC}
\hat{h} = 1, \quad \ddp{\hat{h}}{\hat{x}} = 0 \quad \text{at}~ \hat{x} = 0,
\end{equation}
\begin{equation}\label{E:Modelling:NonDim:FreeBC}
\ddp{^2 \hat{h}}{ \hat{x}^2}  + \poisson \ddp{^2 \hat{h}}{\red{\hat{x}}\blue{\hat{y}}^2} = 0 \quad  \ddp{^3 \hat{h}}{ \hat{x} ^3}  + (2-\poisson) \ddp{^3 \hat{h}}{\hat{x}\partial \y^2}=0 \quad \text{at}~ \hat{x} = 1,
\end{equation}
\begin{align}\label{E:Modelling:NonDim:ContinuityBC1}
\left[ \hat{h} \right]_{-}^{+} = \left[\ddp{\hat{h}}{\hat{x}}\right] _{-}^{+}  = \left[\ddp{\hat{h}}{\hat{y}} \right]_{-}^{+} &= 0 ,\\
 \left[\ddp{^2 \hat{h}}{\hat{x}^2} +\poisson \ddp{^2 \hat{h}}{\hat{y}^2}  \right]_{-}^{+} = \left[\ddp{^2 \hat{h}}{\hat{y}^2} +\poisson \ddp{^2 \hat{h}}{\hat{x}^2}  \right]_{-}^{+}  = \left[\ddp{^2 \hat{h}}{\hat{x} \partial \hat{y}}  \right]_{-}^{+}  &= 0, \label{E:Modelling:NonDim:ContinuityBC2}
\\
\left[\left(\ddp{^3 \hat{h}}{\hat{x}^3} + \ddp{^2 \hat{h}}{\hat{x} \partial \hat{y} ^2}\right) \right]_{-}^{+}  = \left[ \left(\ddp{^2 \hat{h}}{\hat{x}^2 \partial y} + \ddp{^2 \hat{h}}{\hat{y} ^3}\right) \right]_{-}^{+} &=0 \label{E:Modelling:NonDim:ContinuityBC3},
\end{align}
where the jump conditions in~\eqref{E:Modelling:NonDim:ContinuityBC1}--\eqref{E:Modelling:NonDim:ContinuityBC3} are now (and henceforth) evaluated across the dimensionless meniscus position $\hat{x}_m(y,t)$.

The boundary conditions~\eqref{E:Model:Liquid:LaplaceBC} and~\eqref{E:Model:Liquid:nofluxBC} on the pressure become
\abeqn{E:Modelling:NonDim:PressureJumpBC}{
\ddp{\hat{p}}{\hat{x}} = 0 \quad \text{at}~\hat{x} = 0, \qquad \quad \hat{p} = -\bendability\left(\frac{1}{\hat{h}} + \aspect \ddp{^2 \hat{x}_m}{\hat{y}^2}\right)\quad \text{at}~\hat{x} = \hat{x}_m.}
Here, $\aspect = H/(L \cos \theta)$ is a reduced channel aspect ratio, which arises as the ratio between the typical radii of curvature in the transverse and in-plane directions. Note that $\aspect$ always has the same sign as $\bendability$. In particular, this means that their ratio $\bendability / \aspect$, which shall appear frequently, is always positive.

We retain the final term in~\eqref{E:Modelling:NonDim:PressureJumpBC} despite it being higher order in $\aspect \ll 1$ (we assume $\cos \theta \sim \mathcal{O}(1)$); for periodic perturbations with wavenumber $k$, this term is $\mathcal{O}(\aspect k^2)$ and will become important for large wavenumber (short wavelength) perturbations. (Note also that the final term in~\eqref{E:Modelling:NonDim:PressureJumpBC} is larger than the errors introduced by using lubrication theory, which are $\mathcal{O}(\aspect^2)$.)

Finally, the dimensionless meniscus position $\hat{x}_m = \hat{x}_m(\hat{y}, \hat{t})$ evolves according to
\begin{equation}\label{E:Modelling:NonDim:Kinematic}
\ddp{ \hat{x} _m}{\hat{t}} = -\left.\frac{\hat{h}^2}{3|\bendability|}\left(\ddp{\hat{p}}{\hat{x}} - \ddp{\hat{x}_m}{\hat{y}}\ddp{\hat{p}}{\hat{y}} \right)  \right|_{\hat{x} = \hat{x}_m},
\end{equation}
correct to $\mathcal{O}(\aspect^2)$. Henceforth, we drop hats and all variables are assumed dimensionless.

\section{Equilibria}\label{S:Equilibria}
%set out equilibria for both wetting and non-wetting cases
\citet{Taroni2012JFM} considered the $y$-invariant analogue of the system~\eqref{E:Modelling:NonDim:PDE1}--\eqref{E:Modelling:NonDim:Kinematic}, and set out the conditions under which equilibria may exist for wetting conditions ($\bendability > 0$). These equilibria, extended infinitely in the $y$-direction, are also equilibrium configurations in our system, and so we briefly describe them here as a function of dimensionless parameters $\bendability$ (the ability of the liquid to deform the channel walls)  and the liquid volume (defined below) (the channel aspect ratio $a$ enters the perturbation problem described in the following section). This description of equilibria is appropriate also for non-wetting configurations. We also discuss the stability of equilibria to perturbations that are uniform in  the $y$-direction, which facilitates our understanding of the behaviour for perturbations of arbitrary wavelength that follows in \S\ref{S:LSA}. 

We denote the equilibrium meniscus position and channel shape by $x_m = x_0$ and $h = h_e(x)$, respectively, suppressing the $y$-dependence to reflect uniformity in this direction. Following~\citet{Taroni2012JFM}, we use the meniscus position, $x_0$, to parametrize the wall shapes and calculate the corresponding cross-sectional volume,
\begin{equation}\label{E:Equilibria:Volume}
V  = V(x_0) = \int_{0}^{x_0} h_e(x)~\mathrm{d}x.
\end{equation}
\citet{Taroni2012JFM} showed that the equilibria have wall shapes
\begin{equation}\label{E:Equilibria:EqShape}
\h_e(\x) = \left\{\begin{array}{l l}
 \h_0+ \frac{\bendability}{24 \h_0}\left[4 x_0^3( \x_0 - \x) - ( \x_0 - \x)^4\right] &  0 < \x <  \x_0, \\
 \h_0 -\frac{\bendability }{6 h_0}(\x - \x_0) \x_0^3 & \x_0 < \x < 1.
\end{array}\right.
\end{equation}

The uniform pressure associated with~\eqref{E:Equilibria:EqShape} is $p = p_0 = -\bendability/h_0$, where $h_0 \coloneqq  \h_e(\x_0)$. To satisfy the \rout{ pressure boundary condition}\red{~\eqref{E:Modelling:NonDim:PressureJumpBC}b}\blue{clamped condition~\eqref{E:Modelling:NonDim:ClampedBC}} and the volume constraint~\eqref{E:Equilibria:Volume}, $x_0$ and $h_0$ must satisfy
\abeqn{E:Equilibria:MeniscusDispQuadratic}{
\h_0^2 - \h_0 + \frac{\bendability \x_0^4 }{8} = 0, \quad \text{and}\quad V= \x_0 \h_0 + \frac{3\bendability \blue{x_0^5}}{40 \h_0 \red{x_0^5}}, 
}
respectively. In addition, to avoid situations in which the walls touch, we require
\begin{equation}\label{E:Equilibria:NoTouchCond}
\h_e(\x = 1) = \h_0 - \frac{\bendability  (1- \x_0 )\x_0^3}{6 \h_0} > 0.
\end{equation}

Note that equation~\eqref{E:Equilibria:MeniscusDispQuadratic}a has roots
\begin{equation}\label{E:Equilibria:MeniscusWidth}
h_0 = h_0^{\pm} = \frac{1}{2}\left[ 1 \pm \left(1 - \frac{\bendability x_0^4}{2}\right)^{1/2}\right].
\end{equation}
so that at most two equilibria may exist with the same parameter values. For now, we distinguish between the these two possibilities by referring to them as `$+$' and `$-$' roots according to the sign taken in~\eqref{E:Equilibria:MeniscusWidth}; we shall ultimately only be interested in the `$+$' root, the reasons for which are discussed below.

\begin{figure}
\centering
\includegraphics[width =0.98\textwidth]{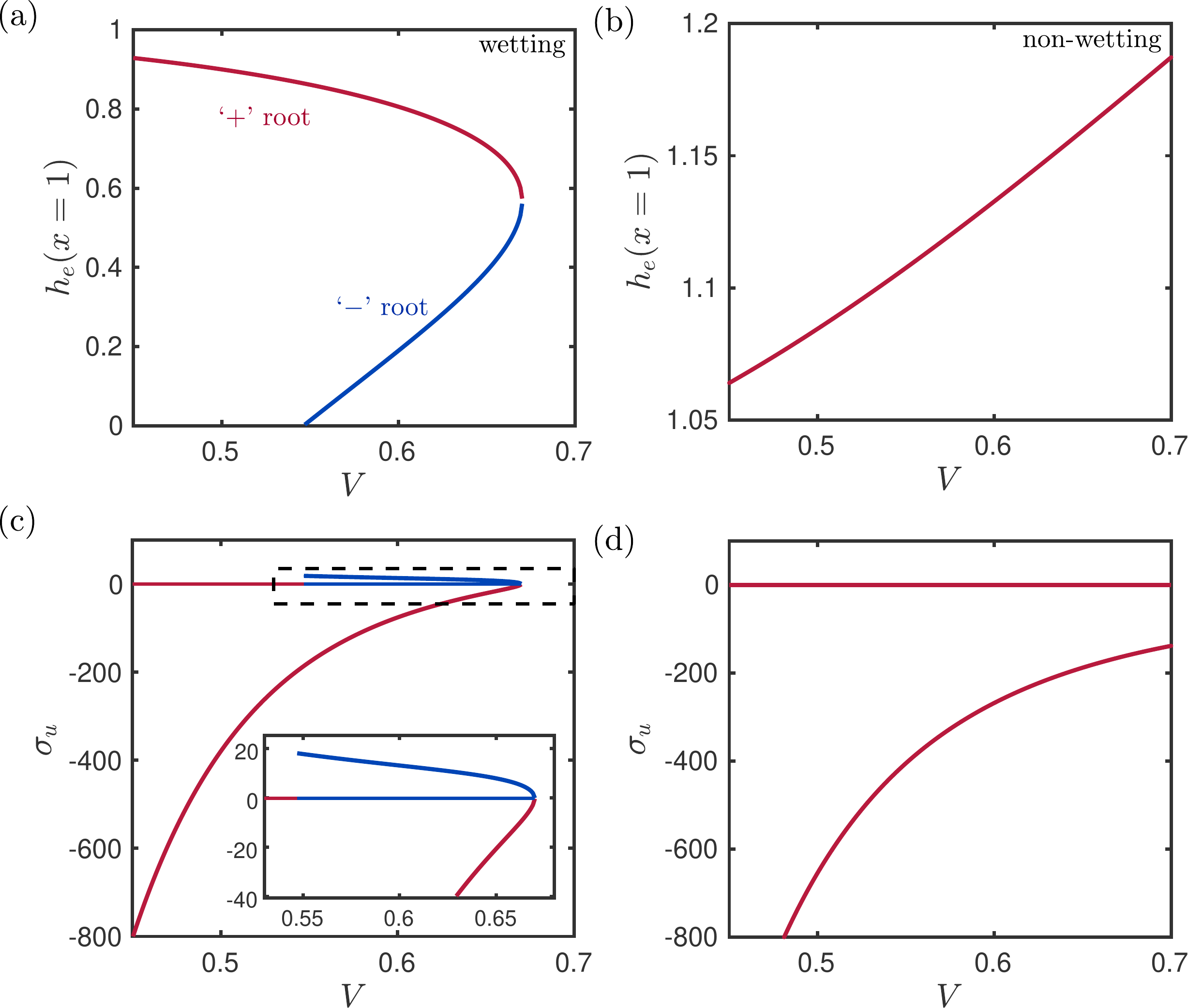}
\caption{(a)--(b) Channel width at the free end, $h_e(x = 1)$, in steady solutions of the model equations~\eqref{E:Modelling:NonDim:PDE1}--\eqref{E:Modelling:NonDim:Kinematic} with (a) $\bendability = 5$ and (b) $\bendability = -5$. Where appropriate, the equilibria corresponding to the `$+$' and `$-$' roots in~\eqref{E:Equilibria:MeniscusWidth} are indicated by red and blue curves, respectively (the latter do not exist in the non-wetting case). (c)--(d) Growth rate $\sigma_u$ of uniform perturbations [i.e.~of the form~\eqref{E:Equilibria:uniform_perturbation}] to the equilibria corresponding to those shown in (a) and (b), respectively. Each equilibria is associated with two values of $\sigma_u$, one of which is always zero. (The red and blue $\sigma_u = 0$ curves are indistinguishable for $0.55 \lesssim V \lesssim 0.67$.) The inset in (c) is a close up of the section of the main figure indicated by the black dashed box.}\label{fig:Equilibria}
\end{figure}

Non-wetting configurations ($\bendability < 0$) always have only a single equilibrium: the `$+$' root of~\eqref{E:Equilibria:MeniscusWidth} exists for all $V$, while the `$-$' root has $h_0 < 0$ for all $V$, which is nonphysical. The channel width at the free end, $h_e(x = 1)$, corresponding to this root, increases monotonically with volume $V$ (figure~\ref{fig:Equilibria}b). %It is not immediately obvious that this should be the case, since increases in volume result in liquid pressure applied over a larger area, which acts to increase the deflection, while the associated reduction in liquid pressure acts in the opposite direction.

In contrast, for wetting configurations ($\bendability >0$), there are regions in which each of none, one, or both of the roots of~\eqref{E:Equilibria:MeniscusWidth} correspond to physically realizable equilibria, depending on the volume $V$ (figure~\ref{fig:Equilibria}a, c). \citet{Taroni2012JFM} described in detail the location of these equilibria; here we simply note that the `$-$' root only exists for a narrow band of sufficiently large $V$, with the no-touching condition~\eqref{E:Equilibria:NoTouchCond} violated at the lower $V$ end of this band (figure~\ref{fig:Equilibria}a), and that the `$+$' root exists for all $V$ up to some finite value (for $\bendability = 5$, this value is approximately 0.68, see figure~\ref{fig:Equilibria}a).

Before moving on to consider the stability of these equilibria to in-plane perturbations, we briefly consider their stability to \textit{uniform} perturbations (or, equivalently to in-plane perturbations with zero wavenumber). In this scenario, the instability mechanism is somewhat simpler than that described in \S\ref{S:Introduction}: the meniscus would like to advance to wet the beams over a longer length, but the additional deformation that results will incur a bending energy penalty.

Following~\citet{Taroni2012JFM}, we probe the stability of the equilibria by substituting
\begin{equation}\label{E:Equilibria:uniform_perturbation}
h = h_e(x) + \varepsilon \Lambda_u(x)\exp(\sigma_u t),\quad p = p_0 + \varepsilon \Pi_u(x)\exp(\sigma_u t),\quad x_0 = x_0 +  \varepsilon \exp(\sigma_u t),
\end{equation}
where $\varepsilon \ll 1$ is arbitrary, into the model equations~\eqref{E:Modelling:NonDim:ClampedBC}--\eqref{E:Modelling:NonDim:Kinematic}; linearizing in $\varepsilon$ yields a boundary value problem (BVP) which can be solved numerically to obtain the growth rate $\sigma_u$ for a given $\bendability$ and $V$; here, and in the following, numerical solutions are obtained using the BVP4c routine implemented in \textsc{matlab}~\citep{Kierzenka2001BVP}. The code used to solve this system of equations is available online~\citep{BendocapillaryRepo}.

For each of the roots of~\eqref{E:Equilibria:MeniscusWidth}, this boundary value problem has two solutions, resulting in two distinct values of $\sigma_u$ for each equilibrium (figure~\ref{fig:Equilibria}c--d). We find that one of these growth rates is always zero, and refers to a uniform advance of the meniscus with no corresponding change in channel shape. This solution corresponds to a situation in which mass is not conserved in sections through the $(x,z)$ plane, but we do not rule out this ostensibly non-physical situation in the following because it is reasonable that liquid might be recruited from the $y$-direction, if variations in this direction are ignored.

The `$-$' root, which is only physically relevant in the wetting case, has a non-zero growth rate that is always positive: these equilibria are unstable to uniform perturbations. In contrast, the `$+$' root, which is physically relevant for both wetting and non-wetting configurations, has a non-zero growth rate that is negative: these equilibria are always stable to uniform perturbations. In addition, for small volumes, these growth rates are large in magnitude. As $V$ increases, the non-zero growth rate associated with the `$+$' root becomes less negative: the bending penalty incurred by the perturbation is reduced when the meniscus is closer to the free end, where the channel is more deformable.

The `$-$' root is always unstable, even without accounting for variations in the in-plane direction, and therefore does not represent a \rout{sensible}\blue{relevant} base state about which to consider the pattern forming instability. In addition, we are primarily interested in small volume configurations; \blue{in particular, any instability arising from the `$-$' root would only be seen when the volume of liquid is sufficiently large (this root only exists for sufficiently large $V$), while an instability arising from the `$+$' root could emerge at any $V > 0$. Therefore, for sufficiently small volume configurations, we would never see the `$-$' root.} \rout{For these two reasons}\blue{Therefore}, we shall, henceforth ignore the equilibria corresponding to the `$-$' root in~\eqref{E:Equilibria:MeniscusWidth}. 

\section{Linear Stability Analysis}\label{S:LSA}
%set out linearized equations and present numerical solutions of them 
In this section, we assess the linear stability of the equilibria described in the previous section to in-plane perturbations with an \textit{arbitrary} wavenumber $k$. We do so by substituting 
\begin{align}
h &= h_e(x) + \varepsilon \Lambda(x)\exp(\sigma t + i k y),\label{E:LSA:Perturbation1} \\
 p &= p_0 + \varepsilon \Pi(x)\exp(\sigma t + i k y),\\ x_m  &= x_0 +  \varepsilon \exp(\sigma t + i k y),\label{E:LSA:Perturbation3}
\end{align}
with arbitrary $\varepsilon \ll 1$, in the model equations~\eqref{E:Modelling:NonDim:ClampedBC}--\eqref{E:Modelling:NonDim:Kinematic}. Linearizing the resulting problem yields the following BVP for channel shape and pressure perturbations $\Lambda$ and $\Pi$, respectively:
\begin{align}
\sigma \Lambda &=  \frac{h_e^2}{3|\bendability|}\left[3\dd{h_e}{x} \dd{\Pi}{x} + h_e\left(\dd{^2 \Pi}{x^2} - k^2 \Pi\right)\right] & &0 < x < x_0,\label{E:LSA:ODEwet}\\
0 &= \Pi & &x_0 < x < 1,\label{E:LSA:ODEdry}\\
\Pi &= \dd{^4 \Lambda}{x^4} - 2k^2 \dd{^2 \Lambda}{x^2} + k^4 \Lambda & & 0 < x <1.\label{E:LSA:pressure2shape}
\end{align}
The boundary conditions on~\eqref{E:LSA:ODEwet}--\eqref{E:LSA:pressure2shape} are
\begin{align}
\Lambda &= 0 = \dd{\Lambda}{x} = \dd{\Pi}{x} & &\text{at}~x = 0,\label{E:LSA:BC_at_0}\\
\Pi &= \frac{\bendability}{\red{h_m}\blue{h_0}^2}\left(\dd{h_e}{x} + \Lambda\right) + \bendability \aspect k^2 & &\text{at}~x = x_0,\label{E:LSA:pressure_bc}\\
\dd{^2 \Lambda}{x^2} - \poisson k^2
\Lambda & = 0, \quad  \dd{^3 \Lambda}{x^3} - (2-\poisson)k^2 \dd{\Lambda}{x} = 0 & &\text{at}~x = 1,\label{E:LSA:BC_at_1}
\end{align}
\begin{equation}\label{E:LSA:jump_conds}
\left[\Lambda\right]_-^+= \left[\dd{\Lambda}{x}\right]_-^+ = \left[\dd{^2\Lambda}{x^2}\right]_-^+= 0, \quad\left[\dd{^3 \Lambda}{x^3}\right]_-^+ = \frac{\bendability}{\red{h_m}\blue{h_0} }.
\end{equation}
\blue{The final equation in~\eqref{E:LSA:jump_conds}
arises from linearizing the no shear boundary condition~\eqref{E:Modelling:NonDim:ContinuityBC3} onto the perturbed interface, thus introducing terms at the next (i.e. fourth) derivative of the base state, which is non-zero. (Recall that the fourth derivative of the channel shape is the pressure, which is inversely proportional to the channel width at the meniscus in the base state.)}
The growth rate $\sigma$ satisfies
\begin{equation}\label{E:LSA:kinematic}
\sigma = -\frac{\red{h_m}\blue{h_0}^2}{3|\bendability|}\left.\ddp{\Pi}{x}\right|_{x = x_0}.
\end{equation}

The terms on the right hand side of~\eqref{E:LSA:pressure_bc} elucidate the mechanism described in \S\ref{S:Introduction}: from left to right, they correspond to the transverse curvature changes arising from channel tapering set by the base state, transverse curvature changes arising from the elastic response to the perturbation, and the stabilizing term that arises from the (locally convex) in-plane curvature of the interface.

\subsection{Numerical Results}

\begin{figure}
\centering
\includegraphics[width =\textwidth]{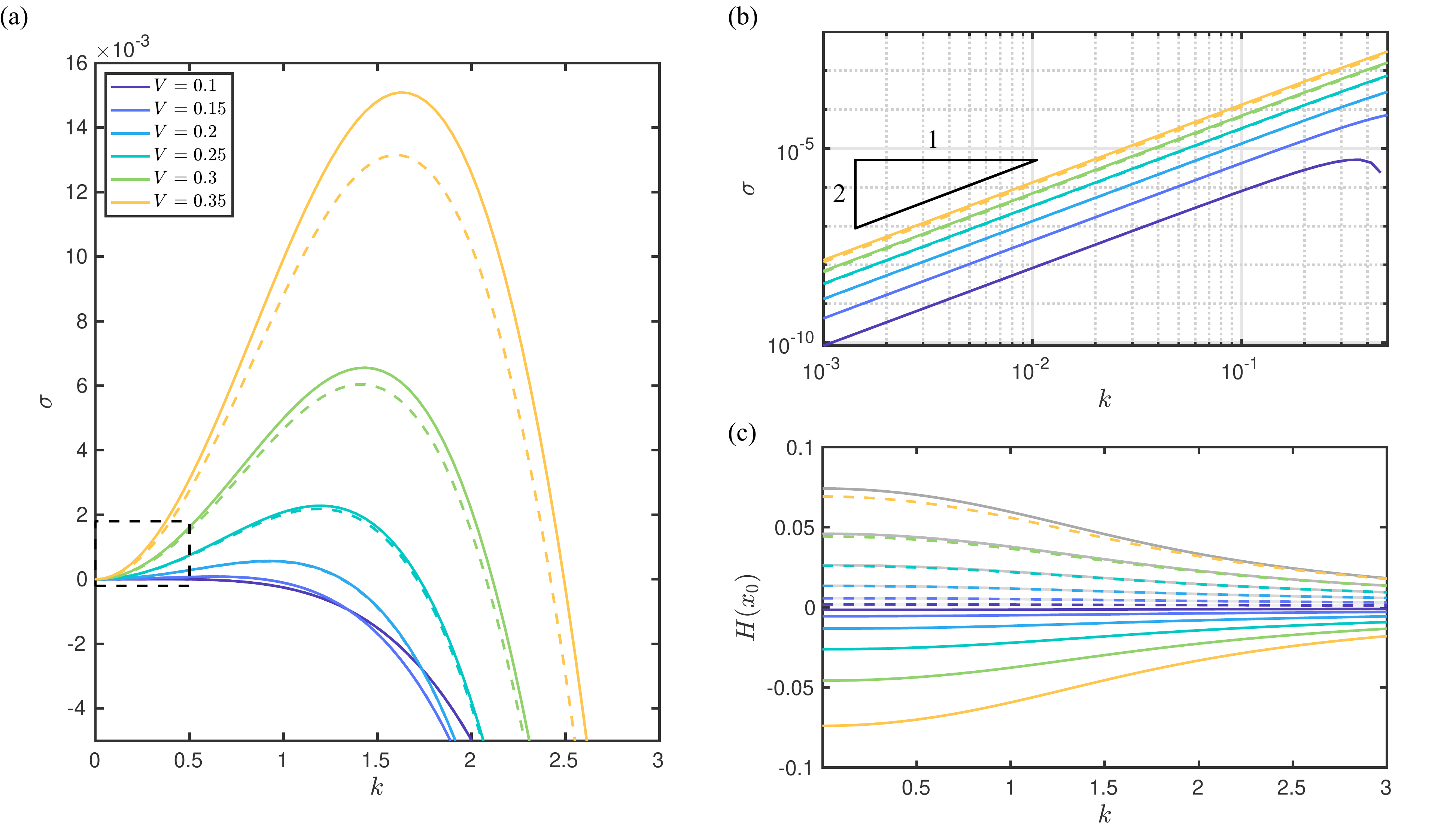}
\caption{(a)--(b) Growth rates $\sigma$, and (c) channel width perturbations, $H(x_0)$, obtained by numerically solving the boundary value problem~\eqref{E:LSA:ODEwet}--\eqref{E:LSA:kinematic} for periodic perturbations with wavenumber $k$ to an equilibrium with cross-sectional volume $V$ (values indicated by the legend). All data correspond to solutions with either $\bendability = 5$, $a = 0.01$ (solid lines) or $\bendability = -5$, $a = -0.01$ (dashed lines). Grey-scale lines in (c) show the reflection of the $\bendability = 5$ curves in the line $H(x_0) = 0$, with darker hues corresponding to larger $V$. The plot (b) is as in (a), but zoomed in on the dashed box in (a), plotted on logarithmic axes. Note that the solid and dashed lines are almost indistinguishable in (b).}
\label{fig:LinearStability:GrowthRates}
\end{figure}
\blue{In this section, we describe numerical solutions of\eqref{E:LSA:ODEwet}--\eqref{E:LSA:kinematic}. As was the case for the uniform perturbation discussed in section~\ref{S:Equilibria}, these solutions are obtained using the BVP4C routine, implemented in \textsc{MATLAB}. In Appendix~\ref{Appendix:Numerics}, we describe this procedure in detail and present convergence tests.}
\red{As was the case for the uniform perturbation discussed in section~\ref{S:Equilibria}}\blue{Again}, we find two distinct numerical solutions of~\eqref{E:LSA:ODEwet}--\eqref{E:LSA:kinematic}, and which of these is returned depends on the proximity to the initial guess for $\sigma$ that is passed to the BVP4c routine. The two branches originate at $k=0$ from the $\sigma_u$ discussed in the previous section. Here we are interested only in the root that originates from the zero solution for $\sigma_u$: this branch shows positive growth rates (figure~\ref{fig:LinearStability:GrowthRates}a), while the branch that originates from the non-zero solution for $\sigma_u$ has very high decay rates even for small $k$~\citep[not shown, but see Ch. 5 of][]{BradleyPhDthesis}. Henceforth, we use $\sigma(k)$ to denote the branch of solutions to~\eqref{E:LSA:ODEwet}--\eqref{E:LSA:kinematic} that originates from the zero solution for $\sigma_u$ (i.e. $\sigma(k=0) = 0$), and ignore the branch that originates from the non-zero solution at $k = 0$.

 %In particular, those solutions on the $\sigma_1$ branch originate from the $k = 0$ solution that does not conserve mass in the $x$-direction; while this is nonphysical for two-dimensional equilibria, in three-dimensions for $k \neq 0$, however, it is possible to move mass in the $y$-direction and there is no contradiction.

The two salient observations from the dispersion relations shown in figure~\ref{fig:LinearStability:GrowthRates} are, firstly, that $\sigma > 0$ for sufficiently small wavenumbers, an observation that appears to be generic in numerical solutions of the BVP~\eqref{E:LSA:ODEwet}--\eqref{E:LSA:kinematic} for both wetting and non-wetting configurations, and, secondly, that $\sigma\sim k^2$ as $k \to 0$ (figure~\ref{fig:LinearStability:GrowthRates}b) as suggested by the scaling argument~\eqref{E:Scaling:amplitude_ode}. We shall derive this formally in the case of small deformations shortly.

Going beyond these observations, we see that numerical solutions also indicate that growth rates are not symmetric in $\bendability \to -\bendability$: growth rates are marginally larger for wetting configurations ($\bendability > 0)$ than non-wetting configurations ($\bendability < 0)$, for the same $|\bendability|$ and $|\aspect|$ (figure~\ref{fig:LinearStability:GrowthRates}a). It is not straightforward to provide a clear link between a reversal of wetting conditions ($\bendability \to -\bendability$, $a \to -a$) and changes in the growth rate, since the pressure gradient and equilibrium meniscus displacement (which set the growth rate via~\eqref{E:LSA:kinematic}) are intimately coupled in the BVP~\eqref{E:LSA:ODEwet}--\eqref{E:LSA:kinematic}. Note, however, that this asymmetry disappears in the limit $V \to 0$ (figure~\ref{fig:LinearStability:GrowthRates}a, b); in the following section, we confirm this symmetry analytically in the case of small deformations (which covers the limit $V \to 0$).  %\blue{I think we should also be able to say qualitatively why the difference disappears as $k \to 0, \infty$, but it doesn't seem immediate to me from~\eqref{E:LSA:ODEwet}--\eqref{E:LSA:kinematic}.}

%discussion of stability and comparison between wetting and non-wetting cases
Another generic feature of numerical solutions of the BVP~\eqref{E:LSA:ODEwet}--\eqref{E:LSA:kinematic} is that $H(x_0)$, the perturbation to the channel width at the meniscus, is negative for wetting configurations and positive for non-wetting configurations (figure~\ref{fig:LinearStability:GrowthRates}c). In other words, the channel deformation of the base state is enhanced at protrusions and reduced at invaginations, for both wetting and non-wetting conditions, confirming the suggestion made when describing the instability mechanism in \S\ref{S:Introduction}. In the results shown in figure~\ref{fig:LinearStability:GrowthRates}c, the channel shape perturbation tends to increase in magnitude with the volume $V$ and decrease with the wavenumber $k$; in the case of small deflections, however, the channel shape perturbation at the meniscus is, perhaps surprisingly, independent of the wavenumber, as we shall see. 

\begin{figure}
\centering
\includegraphics[width = \textwidth]{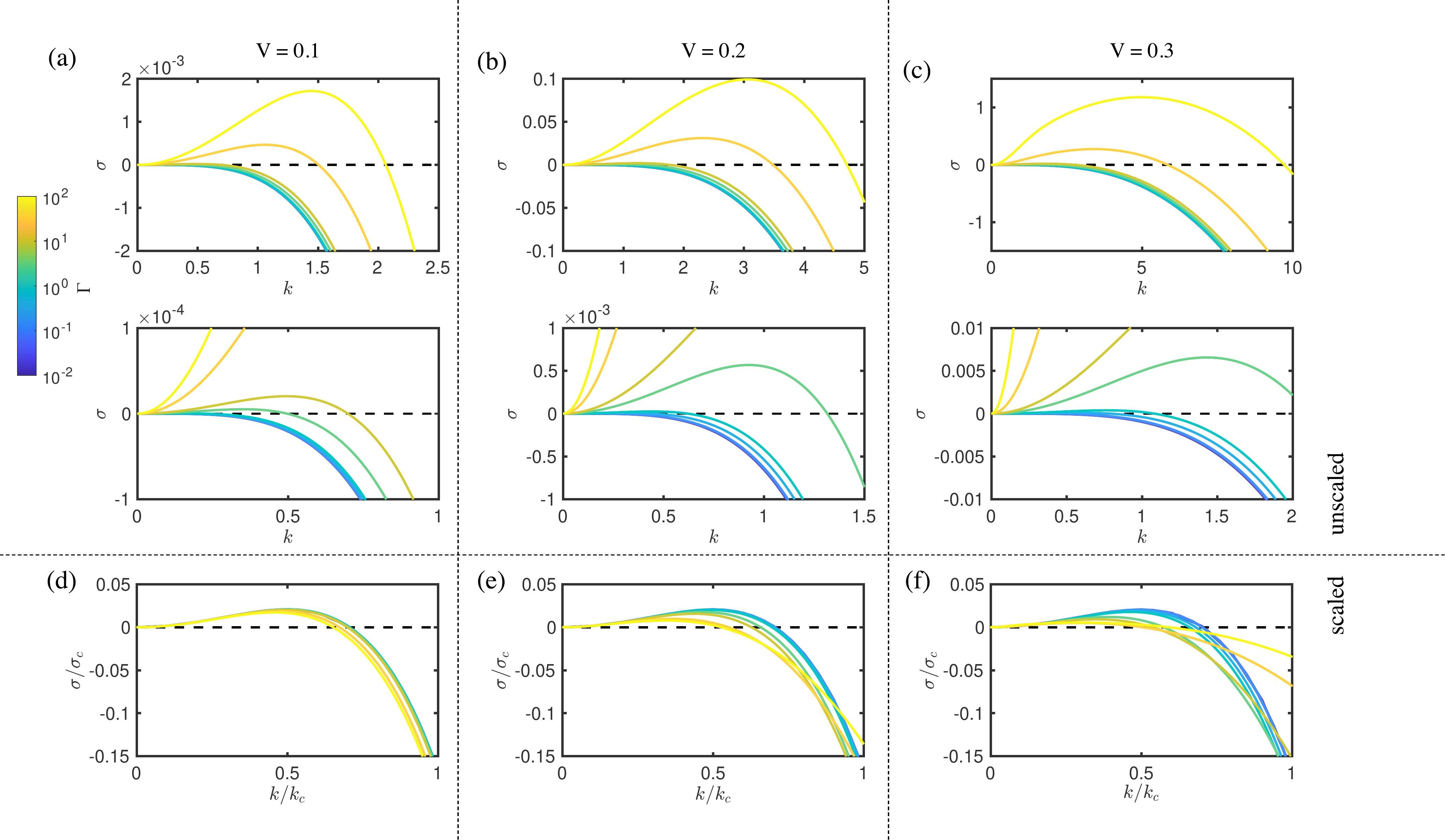}
\caption{Numerically obtained dispersion relations $\sigma(k)$\red{ rescaled according to~\eqref{E:LSA:DimensionlessScalings}} for cross sectional volumes (a) $V = 0.1$, (b) $V = 0.2$, and (c) $V = 0.3$ with $\aspect = 0.01$ in each case. \blue{In each of (a)--(c), the second row is as in the first, but zoomed in around the origin.} \blue{Within each of (a)--(c),} \rout{E}\blue{e}ach curve corresponds to a different value of $\bendability$, taking logarithmically spaced values between $10^{-2}$ and $10^{2}$ as indicated by the colourbar \blue{on the left hand side}\rout{(right)}. Here we show only wetting configurations ($\bendability >0$, $a >0$), but non-wetting configurations behave similarly (see figure~\ref{fig:LinearStability:GrowthRates}). \blue{(d)--(f) Dispersion relations shown in (a)--(c), respectively, rescaled according to~\eqref{E:LSA:DimensionlessScalings}.} }
\label{fig:RescaledGrowthRates}
\end{figure}

For the parameter values used in figure~\ref{fig:LinearStability:GrowthRates}, the fastest growing mode, denoted $k^*$, and the corresponding growth rate $\sigma^* = \sigma(k^*)$, both increase with cross-sectional volume $V$. This is in qualitative agreement with the scaling\blue{s}\rout{ argument}\blue{~\eqref{E:Scaling:CriticalWavenumber} and }~\eqref{E:Scaling:SigmaScaling1} \blue{for the critical wavenumber and growth rate, respectively (as discussed in \S\ref{S:Scaling}).}\rout{, which, after n}\blue{ N}on-dimensionalizing \blue{these} with the length scale $L$ and time scale $\tau_c$,  \rout{suggests the typical scales}\blue{gives a dimensionless critical wavelength and maximum growth rate} 
\begin{equation}\label{E:LSA:DimensionlessScalings}
k_c  =\blue{  \left[\frac{\gamma \cos^2 \theta \red{x_m}\blue{x_0}^3}{B H^3}\right]^{1/2}\times L =} \left(\frac{\bendability V^3}{a}\right)^{1/2}, \qquad \sigma_c = \blue{\left[\frac{\gamma^3 \cos^4 \theta ~ \red{x_m}\blue{x_0}^7}{\mu B^2 H^{4}}\right]\times \tau_c=} \frac{\bendability^2 V^7}{|a|},
\end{equation}
\blue{where the terms in square braces are those identified in~\eqref{E:Scaling:CriticalWavenumber} and~\eqref{E:Scaling:SigmaScaling1}, respectively, with an equilibrium meniscus position denoted by $x_0$.}

\blue{These observations and scaling trends are also borne out by dispersion relations at different values of $\bendability$, which are shown in figures~\ref{fig:RescaledGrowthRates}(a)--(c): for a given volume $V$, solutions with a larger bendability $\bendability$ are associated with a longer critical wavelength and larger maximum growth rate, with a stronger sensitivity to the bendability seen in the critical growth rate than in the critical wavelength.}

\blue{To go beyond these qualitative arguments and make}\rout{For} a quantitative assessment of the scaling argument, we show \blue{in figure~\ref{fig:RescaledGrowthRates}(d)--(f) the same} numerically obtained dispersion relations $\sigma(k)$ \blue{as in figure~\ref{fig:RescaledGrowthRates}(a)--(c)}, rescaled according to~\eqref{E:LSA:DimensionlessScalings}\red{ in figure~\ref{fig:RescaledGrowthRates}a}. Here we see that the data collapse onto a universal curve for $|\bendability| \ll 1$ (indicated by dark shades in figure~\ref{fig:RescaledGrowthRates}), but deviate for $|\bendability|\gg 1$ (light shades). In addition, the deviation occurs sooner (i.e. at lower $\bendability$ values) for larger volume $V$, which correspond to increased base state deformation. To predict the $\sigma/\sigma_c$ curve in the limit $\bendability \to 0$, and elucidate the small deformation results alluded to above, we now present an asymptotic expansion of the eigenvalue problem~\eqref{E:LSA:ODEwet}--\eqref{E:LSA:kinematic} in the limit of small deformations.

\section{Asymptotic Analysis}\label{S:Asymptotics}
\newcommand{\param}{\delta} %small param used in this section
%use asymptotics to describe the envelope of solutions
In this section, we present an asymptotic analysis of the eigenvalue problem~\eqref{E:LSA:ODEwet}--\eqref{E:LSA:kinematic}  in the limit of small channel deformations. The key result of this section is expressing analytically the $|\bendability| \to 0$ limit of the $\sigma/\sigma_c$ curves shown in figure~\ref{fig:RescaledGrowthRates}, thereby predicting the fastest growing mode, and corresponding growth rate, in the limit of small deflections. Along the way, we elucidate the observations mentioned in the previous section for small volumes.

Small deformations are encoded by restricting ourselves to those equilibria with $\epsilon  \coloneqq |\bendability| V^4 \ll 1$. We first note that expanding the volume constraint~\eqref{E:Equilibria:MeniscusDispQuadratic}b gives the following relationship between the volume and meniscus position of the  equilibria:
\begin{equation}\label{E:Asymptotics:EqMeniscusPositionExpansion}
    x_0 = V\left[1 + \mathrm{sgn}(\bendability) \frac{\epsilon}{20} + \order{\epsilon^2}\right],
\end{equation}
in the limit $\epsilon \to 0$. In addition, from~\eqref{E:Equilibria:EqShape}, the equilibrium channel shape is given by
\begin{equation}\label{E:Asymptotics:EqChannelShapeExpansion}
    h_e(x) = 1 + \epsilon \mathrm{sgn}(\bendability)\psi\left(\frac{x}{V}\right) + \order{\epsilon^2}, \quad \text{where} \quad \psi(s) = \frac{1}{24}\left\{\left[4(1-s) - (1-s)^4\right]-3\right\},
\end{equation}
for $x < x_0 = V + \order{\epsilon}$.

Before we can proceed with an asymptotic expansion, we must determine the size of the terms in which the wavenumber $k$ appears. Motivated by the scaling of \S\ref{S:Scaling}, in which the fastest growing modes were those with wavenumbers such that the destabilizing transverse and stabilizing in-plane curvature contributions are comparable (i.e. those values of $k$ that result in the terms on the right-hand side of~\eqref{E:LSA:pressure_bc} being comparable in size), we introduce a scaled wavenumber
\begin{equation}\label{E:Asymptotics:RescaledWavenumber}
k = k_c K = \left(\frac{\bendability V^3}{a}\right)^{1/2} K.
\end{equation}
where $k_c = \sqrt{\bendability V^3 /a}$, as defined in \S\ref{S:Scaling} and~\eqref{E:LSA:DimensionlessScalings}. In addition, in anticipation of the bending deformation being primarily confined to the wet region $0 < x < x_0 = V + \mathcal{O}(\epsilon)$, we introduce the rescaled spatial variable 
\begin{equation}\label{E:Asymptotics:RescaledX}
X = x/V.
\end{equation}

After inserting~\eqref{E:Asymptotics:RescaledWavenumber}\blue{--\eqref{E:Asymptotics:RescaledX}}\rout{and $X$} into the BVP~\eqref{E:LSA:ODEwet}--\eqref{E:LSA:kinematic}, the parameter $\bendability V^5/|a|$ naturally emerges. We write this parameter as the ratio of $\epsilon$ and $\kappa =|a|/V = 2H^2/ (\Omega |\cos \theta| )$, where the latter describes the cross-sectional aspect ratio of the fluid. Our assumption that lubrication theory is applicable requires us to restrict ourselves to $\kappa \ll 1$ (and thus the emergent parameter $ \epsilon / \kappa $ is much larger than $\epsilon$). 

The parameter $\epsilon/\kappa$ can be thought of as capturing the relative sizes of increases of in-plane and transverse bending energies when a small-deformation equilibrium is subject to a perturbation with wavenumber $k \sim k_c$.  To see this, we note that the typical (dimensional) in-plane and transverse wall curvatures induced by this perturbation are
\begin{equation}
\kappa_{\text{in-plane}} \sim k_c^2 \Delta h \sim \frac{\bendability V^3}{a} \Delta h \quad \text{and}\quad \kappa_{\text{transverse}} \sim \frac{\Delta h}{x_0^2} \sim  \frac{\Delta h}{V^2},
\end{equation}
where $\Delta h$ is the typical change in channel thickness that results from the perturbation. The corresponding dimensionless bending energies are
\begin{equation}
E_{\text{in-plane}} \sim \frac{\left(\kappa_{\text{in-plane}}\right)^2}{k_c} \sim \left( \frac{\bendability V^3}{a}\right)^{3/2}(\Delta h)^2,  \quad \text{and}\quad
E_{\text{transverse}} \sim x_0 \left(\kappa_{\text{transverse}}\right)^2 \sim \frac{(\Delta h)^2}{V^3},
\end{equation}
respectively. The ratio of these bending energies is 
\begin{equation}
\frac{E_{\text{in-plane}}}{ E_{\text{transverse}}}
\sim \left(\frac{\epsilon}{\kappa}\right)^{3/2}.
\end{equation}

\renewcommand{\beta}{\mathcal{B}}
To make progress, we consider the limit $\epsilon / \kappa \to 0$, which is possible with $\epsilon/\kappa \gg \epsilon$ provided $\epsilon \lll 1$, and corresponds to small in-plane bending deformations compared to transverse bending deformations (as was imposed explicitly in the scaling argument of \S\ref{S:Scaling}). For a full treatment of the asymptotic problem, one should pose a bivariate asymptotic expansion of each of the perturbation to the channel shape, $H$, the perturbation to the pressure,
$P$, and the growth rate, $\sigma$, in both $\epsilon$ and $\kappa$. In doing so, however, it is seen that the expression for the $\sigma/\sigma_c$ curves in the limit $\bendability \to 0$, as well as the leading ($\order{1}$) and first ($\order{\kappa}$) problems that emerge from such an asymptotic expansion, are independent of the choice of relationship between $\epsilon$ and $\kappa$ \cite[see][which contains a complete treatment of this problem]{BradleyPhDthesis}. For simplicity, therefore, we present here only the case of a distinct limit with $\epsilon \sim \kappa^2$; to reflect this, we introduce the parameter $\beta = \epsilon/\kappa^2$ which is assumed to be $\order{1}$.

Before we proceed, it is instructive to introduce a rescaled channel shape perturbation $H$ and pressure perturbation $P$:
\begin{equation}\label{E:Asymptotics:Rescaling}
G(X) = \frac{H(X)}{\bendability V^3}, \qquad Q(X) = \frac{P(X)}{\bendability V^3}.
\end{equation}
These scalings reflect the leading order behaviour of the perturbation: a shear force of magnitude $\bendability$ applied over a length equal to  the magnitude  of  the  perturbation  (the  shear  is  the  third  derivative  of  the  channel deformation, which, when combined with the length rescaling, is responsible for the $V^3$). 

After inserting~\eqref{E:Asymptotics:EqMeniscusPositionExpansion}, \eqref{E:Asymptotics:EqChannelShapeExpansion}, and~\eqref{E:Asymptotics:Rescaling} into the linearized problem~\eqref{E:LSA:ODEwet}--\eqref{E:LSA:kinematic}, the problem for the $G$, $Q$, and the growth rate $\sigma$ reads, correct to $\mathcal{O}(\epsilon^2)$:
\begin{align}
3\epsilon V^2 \sigma G &= \left(1 + 2\epsilon \psi\right)\left[\epsilon\dd{\psi}{X} \dd{Q}{X} + \left(1 + \epsilon \psi\right)\left(\dd{^2 Q}{X^2} - \beta \epsilon^{1/2} K^2 Q\right)\right] & &0 < X < 1,\label{E:Asymptotics:ODEwet}\\
0 &= Q & &1 < X < \frac{1}{V},\label{E:Asymptotics:ODEdry}\\
Q &= \dd{^4 G}{X^4} - 2\beta \epsilon^{1/2} K^2 \dd{^2 G}{X^2} +\beta^2 \epsilon K^4 G & &0 < X<\frac{1}{V},\label{A:Asymptotics:state_BVP:pressure2shape}
\end{align}
with boundary conditions
\begin{align}
G &= 0 = \dd{G}{X} = \dd{Q}{X} & &\text{at}~X = 0,\label{E:Asymptotics:BC_at_0}\\
Q + \frac{\epsilon V}{20}\ddp{Q}{X} &= \epsilon\left(\dd{\psi}{X} + G + K^2\right)  & &\text{at}~X= 1,\label{E:Asymptotics:pressure_bc}\\
\dd{^2 G}{X^2} -\beta \epsilon^{1/2} \poisson K^2 G &= 0, \quad  \dd{^3 G}{X^3} - (2-\poisson)\beta \epsilon^{1/2} K^2 \dd{G}{X} = 0 & &\text{at}~X = \frac{1}{V},\label{E:Asymptotics:BC_at_1}
\end{align}
\begin{align}\label{E:Asymptotics:jump_conds}
\left[G\right]_-^+= \left[\dd{G}{x}\right]_-^+ = \left[\dd{^2G}{X^2} + \frac{\epsilon V}{20}\dd{^3 G}{X^3}\right]_-^+&= 0, \\
\left[\dd{^3 G}{X^3} + \frac{\epsilon V}{20}\dd{^4 G}{X^4}\right]_-^+ &= 1 - \epsilon \psi(1).
\end{align}
Here the jump applies at the linearized equilibrium position, $X = 1$. The growth rate $\sigma$ satisfies
\begin{equation}\label{E:Asymptotics:kinematic}
3V^2\sigma = -\left[1 + 2\epsilon \psi(1)\right]\left[\dd{Q}{X} + \frac{\epsilon V}{20}\dd{^2 Q}{X^2}\right]_{X = 1}.
\end{equation}
Note that the reduced problem~\eqref{E:Asymptotics:ODEwet}--\eqref{E:Asymptotics:kinematic} is independent of the sign of $\bendability$ and $\aspect$, demonstrating that the growth rate $\sigma$ is independent of the wettability in the limit of small deformations, as suggested in \S\ref{S:LSA}.

To proceed, we pose an asymptotic expansion in powers of $\epsilon^{1/2}$:
\begin{align}
G(X) &=    G_0(X) + \epsilon^{1/2} G_1(X) + \epsilon G_2(X)+ \dots,\label{E:Asymptotics:ExpansionG} \\
Q(X) &=  Q_0(X) + \epsilon^{1/2} Q_1(X) + \epsilon Q_2(X)+\dots,\label{E:Asymptotics:ExpansionQ} \\
\sigma(k) &= \sigma_0 + \epsilon^{1/2} \sigma_1 + \epsilon \sigma_2+ \dots. \label{E:Asymptotics:ExpansionSigma}
\end{align}

In Appendix~\ref{A:SmallDeformationAsymptotics}, we set out the hierarchy of problems that emerge from inserting~\eqref{E:Asymptotics:ExpansionG}--\eqref{E:Asymptotics:ExpansionSigma} into~\eqref{E:Asymptotics:ODEwet}--\eqref{E:Asymptotics:kinematic}, as well as their solution.  We find that the leading and first order pressure profiles $Q_i(X),~i = 1,2$ are linear functions of $X$. However, to satisfy the no-flux condition at $X =0$ [the third of~\eqref{E:Asymptotics:BC_at_0}], this pressure perturbation is in fact constant, and thus from~\eqref{E:Asymptotics:kinematic} offers no contribution to $\sigma$, i.e.
\begin{equation}
\sigma_{0} = \sigma_{1} = 0.
\end{equation}
The leading order contribution to the perturbation to the channel shape is
\begin{equation}\label{E:Asymptotics:ChannelShapeSolution}
G_{0} =\begin{cases}
\frac{-X^3}{6} & 0 < X < 1,\\
\frac{-1}{6}(3X-2) & 1 < X < 1/V.
\end{cases}
\end{equation}

\begin{figure}
\centering
\includegraphics[width = \textwidth]{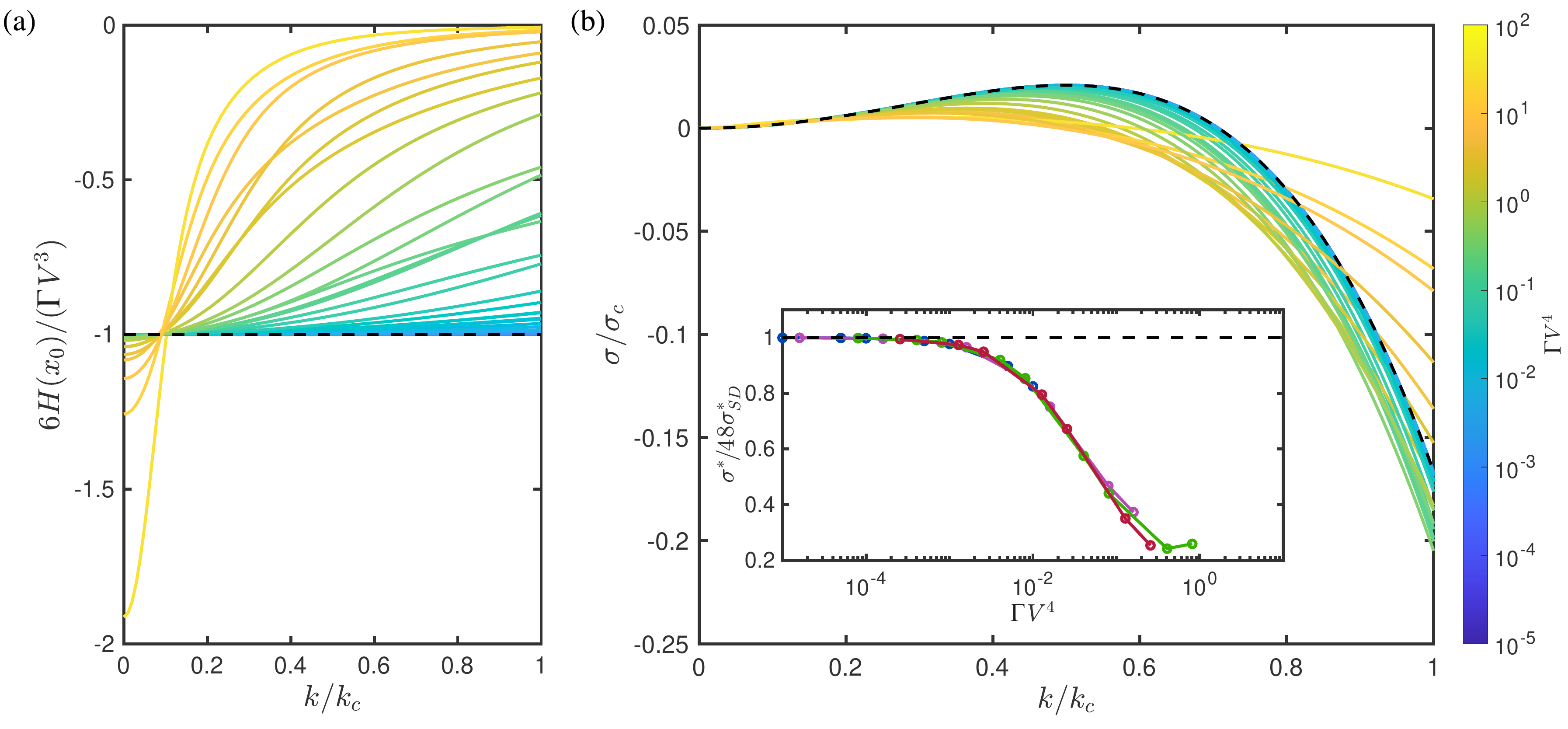}
\caption{(a) Numerically obtained values for the normalized perturbation to the channel width $H(x = x_0)$ and (b) normalized growth rate $\sigma$ as a function of the reduced wavenumber $k/k_c$. Each curve corresponds to a unique $(V,\bendability)$ pair (the aspect ratio $a = 0.01$ is fixed), whose combination $\epsilon = \bendability V^4$ is indicated by the colours in the colourbar. The black dashed curves in (a) and (b) correspond to the asymptotic results~\eqref{E:Asymptotics:ChannelShapeSolution} and~\eqref{E:Asymptotics:FastestGrowingMode}, respectively. The numerical results are indistinguishable from the asymptotic curves for $\epsilon \lesssim 10^{-2}$. The inset in (b) is a semilogarithmic plot of the numerically obtained values of the maximum growth rate $\sigma^*$, rescaled according to~\eqref{E:Asymptotics:FastestGrowingMode}, for $V = 0.1$ (blue), $0.2$ (pink), $0.3$ (green), and $0.4$ (red). [These curves are almost indistinguishable, and terminate where the corresponding equilibria cease to exist, having violated the no contact condition~\eqref{E:Equilibria:NoTouchCond}.] The black dashed curve indicates the small-deformation prediction $\sigma^* = 48\sigma^*_{SD}$ [equation~\eqref{E:Asymptotics:FastestGrowingMode}].}
\label{fig:CollapsedGrowthRates}
\end{figure}

In particular, this means that $H(X=1) = \bendability V^3 G(X=1)\sim -\bendability V^3/6$ as $\epsilon = |\bendability| V^4 \to 0$. This result agrees well with numerical solutions of the BVP~\eqref{E:LSA:ODEwet}--\eqref{E:LSA:kinematic} (figure~\ref{fig:CollapsedGrowthRates}a). Physically, this confirms that the perturbation to the channel shape is negative (positive, respectively) for wetting (non-wetting) configurations, and that the leading order perturbation to the channel shape is independent of the wavenumber $k$, as suggested in \S\ref{S:LSA}. In addition, we find that the Poisson's ratio $\poisson$ does not enter the leading order solution, but does appear in the first order term, $G_{1}$; this indicates that the contribution to the problem from the dry region enters at lower order (the Poisson's ratio only enters the problem via the boundary conditions on the dry region, at $x = 1$), i.e. bending deformations are confined primarily to the wet region in the limit of small, transverse-direction dominated, deformations, as expected.

Most importantly, we find that the first non-zero term in the expansion of $\sigma$~\eqref{E:Asymptotics:ExpansionSigma} is
\begin{equation}\label{E:Asymptotics:LeadingOrderSigma}
\sigma_{3} = \frac{K^2\left(1-2K^2\right)}{6V^2},
\end{equation}
[It might be expected that, because the inhomogeneous right-hand side of~\eqref{E:Asymptotics:pressure_bc} is $\order{\epsilon}$, the first non-zero term in~\eqref{E:Asymptotics:ExpansionSigma} would be $\sigma_2$. However, although 
the first non-zero term in the expansion~\eqref{E:Asymptotics:ExpansionQ} is $Q_{2}$, we find that $Q_{2}$ is in fact constant; the first term in~\eqref{E:Asymptotics:ExpansionQ} with a non-zero gradient, which sets the growth rate, comes in at the next order, $\mathcal{O}(\epsilon^{3/2})$.]

Noting that $\sigma_c = \epsilon^2/\kappa V^2$ in the notation of this section, substituting~\eqref{E:Asymptotics:LeadingOrderSigma} in to the expansion~\eqref{E:Asymptotics:ExpansionSigma}, gives
\begin{equation}\label{E:Asymptotics:SigmaResult}
\frac{\sigma(K)}{\sigma_c} \sim  \frac{K^2\left(1-2K^2 \right)}{6}
\end{equation}
as $\epsilon \to 0$. The right-hand side of~\eqref{E:Asymptotics:SigmaResult} gives a limiting curve that can be compared with numerical solutions for $\sigma(k)/\sigma_c$ (figure~\ref{fig:CollapsedGrowthRates}b). As expected from~\eqref{E:Asymptotics:ExpansionSigma}, numerical solutions with larger values of $\epsilon = \bendability V^3$ deviate more significantly from this limiting curve.

By maximizing~\eqref{E:Asymptotics:SigmaResult} with respect to $K$, we find that the small deformation estimates of the fastest growing mode, denoted $k^*_{SD}$, and the corresponding growth rate, denoted $\sigma^*_{SD}$, are
\begin{equation}\label{E:Asymptotics:FastestGrowingMode}
\sigma^*_{SD}= \frac{1}{48}\sigma_c = \frac{1}{48}\frac{\bendability^2 V^7}{|\aspect|}, \qquad k^*_{SD}= \frac{1}{2}k_c = \frac{1}{2}\sqrt{\frac{\bendability V^3}{\aspect}}.
\end{equation}
The small-deformation predictions~\eqref{E:Asymptotics:FastestGrowingMode} agree well with numerical solutions of the eigenvalue problem~\eqref{E:LSA:ODEwet}--\eqref{E:LSA:kinematic} (see inset of figure~\ref{fig:CollapsedGrowthRates}b, in which perfect agreement would correspond to a constant value of unity in the abscissa). It is interesting to note that the asymptotic result~\eqref{E:Asymptotics:FastestGrowingMode} overestimates the fastest growing mode as $\epsilon$ grows from zero; briefly, this is because the increased penalty from in-plane bending as $V$ increases and base-state deformations grow (which suppresses the growth rate relative to the asymptotic result) is stronger than the effect of increased deformation of the base state (because the meniscus pressure scales with $1/h$, the narrower the confinement at the meniscus, the greater the change in pressure there when the meniscus is perturbed).

\section{Discussion}\label{S:Conclusion}
%very broad description
In this paper, we set out to understand\blue{ a novel bendocapillary instability that is driven by the competition between interfacial curvatures at the liquid surface, and is mediated by the elasticity of the channel in response to liquid pressure. Such interactions are a fundamental component of} the periodic pattern that is observed in experiments in which liquid condenses slowly into deformable microchannels. \rout{In doing so, we identified a novel bendocapillary instability mechanism that is driven by the competition between interfacial curvatures at the liquid surface, and is mediated by the elasticity of the channel in response to liquid pressure. This}\blue{The} bendocapillary instability\blue{ introduced here} is theoretically possible in the same channel for both wetting and non-wetting liquids, \rout{unlike the apparently analogous instability in rigid, tapered channels}\blue{which is not the case for the similar instabilities in rigid, tapered channels that were described by}~\citet{AlHousseiny2012NaturePhysics} \blue{and~\citet{Keiser2016JFM}}.

%what did we do: studied system which may be susceptible, scaling argument, develop mathematical model, linear stability analysis, asymptotics (and why did we do these
We developed a mathematical model of this mechanism, which was simplified by exploiting the small aspect ratio of the both the channel walls and the cavity between them, allowing us to appeal to linear plate theory to describe deformations of the channel walls and lubrication theory to describe the flow. In non-dimensionalizing this system, we identified three dimensionless parameters, relating to the ability of the liquid to deform the channel walls ($\bendability$), the liquid volume ($V$), and the channel aspect ratio ($a$). Equilibrium configurations, which form the base states of the system, are parametrized by the first two of these. 

The rest of the paper focused on a study of the linear stability of these equilibria to in-plane perturbations. We formulated the linearized equations that must be satisfied by perturbations; these equations illustrate explicitly the two ways that the elastic case differs from the tapered, rigid case: the bulk channel deformation is set by the liquid pressure, and the perturbation induces an additional elastic response of the channel. Numerical solutions of the linearized equations suggested three key results, which were verified analytically in the limit of small deformations: (i) both wetting and non-wetting equilibria are always unstable to perturbations of a sufficiently small wavenumber, (ii) the growth rate of the fastest growing mode is highly sensitive to the amount of liquid within the channel ($\sigma \sim V^7$), and (iii) the additional elastic response to the perturbation always enhances the destabilizing transverse curvature contribution, and thus tends to promote instability.

%% above paragraph replaced below %%
%By performing a linear stability analysis of its equilibrium configurations, we identified that the system is unstable to perturbations of a sufficiently small wavenumber. The linearized equations elucidate the two main ways that the elastic case differs from the tapered, rigid case in two important ways: the bulk channel elasticity is set by the liquid pressure, and the channel responds to the perturbation in a way that tends to enhance the difference between the in-plane and transverse curvatures, increasing the range of unstable wavenumbers. The growth rate of the fastest growing mode is highly sensitive to the amount of liquid within the channel (parametrized by the cross-sectional volume $V$); in particular, we identified that the growth rate $\sigma \sim V^7$ in the case of small deformations and negligible in-plane bending from both a simple scaling argument and a formal asymptotic analysis.

%but we only looked at constant volume cases, introduce changing volume 
%In this first part one of two, we considered only the no condensation case, in which equilibria have a constant cross sectional volume $V$; in the second part, we describe how a non-zero condensation rate (i.e. how changing the volume of liquid in the channel) changes the picture presented here. 

The sensitive dependence on the amount of liquid in the channel suggests that the results of this paper may be significantly different when a non-zero condensation rate is included. In the motivating experiments, however, condensation is very slow and an order of magnitude estimate for the experimentally observed wavelength may therefore be obtained from the analysis presented here. Using values from~\cite{Seemann2011JPhysCondMat}, we find that $\bendability \approx -12$ and $\aspect \approx -0.38$ in these experiments; assuming a channel that is half filled with liquid ($V = 0.5$), the asymptotic result~\eqref{E:Asymptotics:FastestGrowingMode} predicts a wavelength of $370~\si{\micro \meter}$, which agrees in its order of magnitude with the wavelength of approximately $200~\si{\micro \meter}$ that is observed experimentally (see figure~\ref{fig:Experiments}). (Smaller values of $V$ result in predictions of longer wavelengths, but these are of the same order of magnitude, for example the scaling argument predicts a wavelength of $4~\si{\milli \meter}$ with $V = 0.1$.) \blue{This agreement in the order of magnitude provides evidence that the instability seen in the experiments results from bendocapillary interactions. However, we stress that the modelling results presented here are not intended to be immediately applicable to the experimental system because of the number of important facets of this system we have neglected to include in our model, such as contact line dynamics and the complex geometry. Indeed, any attempt to develop a quasi two-dimensional experiment may find that instability is nucleated at the edge of the system, rather than within the bulk, as assumed here.}

%our results suggests that configurations are always unstable to \emph{some} wavelength, but we don't expect to see instability in practice because of (and not limited to) (a) slow growth rates (b) finite size effects and (c) contact angle hysteresis
As mentioned, our results suggest that liquid sitting in narrow, deformable channels on small scales are always unstable to perturbations of sufficiently long wavelengths. There are several reasons why we do not expect this to be the case in practice. Firstly, realistic channels will have a finite extent in the $y$-direction (as it is referred to here) and the maximum wavelength of perturbations will be restricted to this length. \rout{Secondly}\blue{Moreover}, configurations with stiff walls (small $\bendability$), will have fastest growing modes whose growth rates are very small [the growth rate $\sigma \sim \bendability^2$ see equation~\eqref{E:Asymptotics:FastestGrowingMode}], allowing processes that occur on longer timescales (e.g. evaporation or condensation) to interact with the growth of the bendocapillary instability. Finally, we made a series of assumptions on the physical processes included in our model. Perhaps most notably, we have neglected dynamic contact angle effects, which have been shown to be important in controlling the dynamic behaviour in similar bendocapillary systems~\citep[e.g][]{Bradley2021PRF}. Contact angle hysteresis is expected to reduce the growth rate of perturbations: perturbations that are protrusions, which are advancing interfaces, will have a higher contact angle, and thus smaller magnitude pressure, than invaginations, which are receding interfaces.

%implications: can this instability be exploited
We postulate that the fact this bendocapillary instability has not been described in detail previously might suggest it has not been encountered in any situations in which it is a hindrance. We hope, therefore, that the insights offered in this paper might motivate further experimental and theoretical studies to quantify, and understand, such bendocapillary instabilities and identify situations in which they might be exploited. \\

This publication is based in part upon work supported by the European Research Council (ERC) under the European Union’s Horizon 2020 research and innovation program (Grant agreement No. 637334, GADGET to D.V.), and the Leverhulme Trust (D.V.). \\ 

Declaration of Interests: The authors report no conflict of interest.

Acknowledgements: We are grateful to Martin Brinkmann and Ralf Seemann for the providing the experimental images shown in figure~\ref{fig:Experiments} and for valuable discussions, which improved both descriptions of experiments and their interpretation.

\appendix
\section{Asymptotic Analysis of the Small Deformation Problem}\label{A:SmallDeformationAsymptotics}
In this appendix, we present an asymptotic analysis of the eigenvalue problem~\eqref{E:Asymptotics:ODEwet}--\eqref{E:Asymptotics:kinematic}, which emerges as a rescaled form of the full problem~\eqref{E:LSA:ODEwet}--\eqref{E:LSA:kinematic}, in the case that $\epsilon \sim \kappa^2$. Recall that to make progress, we pose an asymptotic expansion in powers of $\epsilon^{1/2}$:
\begin{align}
G(X) &=    G_0(X) + \epsilon^{1/2} G_1(X) + \epsilon G_2(X)+ \dots, \label{A:E:G_expansion}\\
Q(X) &=  Q_0(X) + \epsilon^{1/2} Q_1(X) + \epsilon Q_2(X)+\dots,\label{A:E:Q_expansion}\\
\sigma(k) &= \sigma_0 + \epsilon^{1/2} \sigma_1 + \epsilon \sigma_2+ \dots. \label{A:E:sigma_expansion}
\end{align}

After introducing~\eqref{A:E:G_expansion}--\eqref{A:E:sigma_expansion} into~\eqref{E:Asymptotics:ODEwet}--\eqref{E:Asymptotics:kinematic}, we obtain a hierarchy of problems by balancing powers of $\epsilon^{1/2}$. The leading order [$\order{1}$] problem reads
\begin{align}
0&= \dd{^2Q_0}{X^2} & &0 < X<1,\label{A:Asymptotics:0thOrder:ODEwet}\\
0&= Q_0  & &1 < X< \frac{1}{V},\\
Q_0 &= \dd{^4 G_0}{X^4} & & 0 < X < \frac{1}{V},
\end{align}
\begin{align}
G_0 &= 0 = \dd{G_0}{X}= \dd{Q_0}{X} & &\text{at}~X = 0,\label{A:Asymptotics:0thOrder:x0bc}\\
Q_0 &=0 & &\text{at}~X = 1,\label{A:Asymptotics:0thOrder:pressurebc}\\
\dd{^2 G_0}{X^2}&=0, \quad  \dd{^3 G_0}{X^3} = 0 & &\text{at}~X = \frac{1}{V},
\end{align}
\begin{equation}
\left[G_0\right]_-^+ =0 = \left[\dd{G_0}{X}\right]_-^+ = \left[\dd{^2 G_0}{X^2}\right]_-^+, \quad  \left[\dd{^3 G_0}{X^3}\right]_-^+ = 1,
\end{equation}
\begin{equation}\label{A:Asymptotics:0thOrder:kinematic}
3V^2 \sigma_0 = -\left.\dd{Q_0}{X}\right|_{X=1}.
\end{equation}
From~\eqref{A:Asymptotics:0thOrder:ODEwet}, we see that the pressure $Q_0$ is a linear function of $X$. However, from~\eqref{A:Asymptotics:0thOrder:x0bc} and \eqref{A:Asymptotics:0thOrder:pressurebc}, this linear function has no slope and passes through zero, i.e. $Q_0 = 0$. Then, from~\eqref{A:Asymptotics:0thOrder:kinematic}, we have $\sigma_0 = 0$. The solution to~\eqref{A:Asymptotics:0thOrder:ODEwet}--\eqref{A:Asymptotics:0thOrder:kinematic} for the channel shape is
\begin{equation}\label{A:Asymptotics:0thOrder:solution}
G_0  =\begin{cases}
\frac{-X^3}{6} & 0 < X < 1,\\
\frac{-1}{6}(3X-2) & 1 < X < 1/V.
\end{cases}
\end{equation}

The first order [$\order{\epsilon^{1/2}}$] problem is similar, and the equations for $Q_1$ are identical to those for $Q_0$; we again get no pressure contribution, $Q_1 = 0$ and thus $\sigma_1 = 0$. The shape contribution $G_1$ is non-trivial, but is not required for the determination of the leading order behaviour for $\sigma$, and we therefore do not state it here. We note, however, that the Poisson's ratio $\poisson$ first appears in this term, highlighting the lower order contribution of the dry regions, as mentioned in the main text.

%%%%%%%%%%%%%%%%%%%%%%%%%
The  $\mathcal{O}(\epsilon)$ problem is expressed more simply be exploiting the $\mathcal{O}(1)$ and $\mathcal{O}(\epsilon^{1/2})$ problems. We give only this simplified form here:
\begin{align}
 \dd{^2 Q_2}{X^2}&=0 & &0 < X<1,\\
Q_2 &= 0 & &1 < X< \frac{1}{V},\\
Q_2 &= \dd{^4 G_2}{X^4} - 2K^2\dd{G_1}{X^2}+ K^4 G_0 & & 0 < X < \frac{1}{V},
\end{align}
\begin{align}
G_2 &= 0 = \dd{G_2}{X}= \dd{Q_2}{X} & &\text{at}~X = 0,\\
Q_2 &=G_0+ K^2+ \dd{\psi}{X} & &\text{at}~X = 1,\label{A:Asymptotics:2ndOrder:PressureBC}\\
\dd{^2 G_2}{X^2} - \poisson K^2 G_1 &=0 = \dd{^3 G_0}{X^3}  - (2- \poisson)K^2 \dd{G_1}{X}& &\text{at}~X = \frac{1}{V},
\end{align}
\begin{equation}
\left[G_2\right]_-^+ =0 = \left[\dd{G_2}{X}\right]_-^+ = \left[\dd{^2 G_2}{X^2} + \frac{\beta V}{20}\dd{^3G_0}{X^3}\right]_-^+ = \left[\dd{^3 G_2}{X^3} + \frac{\beta V}{20}\dd{^4 G_0}{X^4}\right]_-^+,
\end{equation}
\begin{equation}
3V^2 \sigma_2 = -\left.\dd{Q_2}{X}\right|_{X=1}.
\end{equation}
Crucially, the boundary condition~\eqref{A:Asymptotics:2ndOrder:PressureBC} is inhomogeneous, in contrast to the corresponding boundary condition for the lower order problems [e.g.~\eqref{A:Asymptotics:0thOrder:pressurebc}]. We therefore find the first non-zero pressure term in the expansion~\eqref{E:Asymptotics:ExpansionQ} for the channel shape perturbation to be
\begin{equation}\label{A:Asymptotics:2ndOrder:solQ2}
Q_2 = \begin{cases}
K^2 - \frac{1}{2} & 0 <X  < 1,\\
0 & 1 < X < 1/V,
\end{cases}
\end{equation}
where we have used $G_0$, from~\eqref{A:Asymptotics:0thOrder:solution}, to obtain $Q_2$. This leading order pressure contribution is constant in the liquid and thus again offers no contribution to the growth rate, hence $\sigma_2 = 0$.

To obtain a non-zero term in the expansion for $\sigma$ we must proceed to $\mathcal{O}(\epsilon^{3/2})$, where we find that
\begin{align}
 \dd{^2 Q_3}{X^2} - K^2 Q_2 &= 0 & &0 < X<1,\label{A:Asymptotics:3ndOrder:odewet}\\
Q_3 &= 0 & &1 < X< \frac{1}{V},\\
Q_3 &= \dd{^4 G_3}{X^4} - 2K^2\dd{G_2}{X^2}+ K^4 G_1 & & 0 < X < \frac{1}{V},
\end{align}
\begin{align}
G_3 &= 0 = \dd{G_3}{X}= \dd{Q_3}{X} & &\text{at}~X = 0,\label{A:Asymptotics:3ndOrder:x0bc}\\
Q_3 &=\beta G_1 & &\text{at}~X = 1,\label{A:Asymptotics:3ndOrder:PressureBC}\\
\dd{^2 G_2}{X^2} - \poisson K^2 G_1 &=0 = \dd{^3 G_0}{X^3}  - (2- \poisson)K^2 \dd{G_1}{X}& &\text{at}~X = \frac{1}{V},
\end{align}
\begin{equation}
\left[G_3\right]_-^+ =0 = \left[\dd{G_3}{X}\right]_-^+ = \left[\dd{^2 G_3}{X^2} + \frac{\beta V}{20}\dd{^3G_1}{X^3}\right]_-^+ = \left[\dd{^3 G_3}{X^3} + \frac{\beta V}{20}\dd{^4 G_1}{X^4}\right]_-^+,
\end{equation}
\begin{equation}\label{A:Asymptotics:3ndOrder:kinematic}
3V^2 \sigma_3 = -\left.\dd{Q_3}{X}\right|_{X=1}.
\end{equation}
From~\eqref{A:Asymptotics:3ndOrder:odewet} and~\eqref{A:Asymptotics:3ndOrder:x0bc}, we find that
\begin{equation}\label{A:Asymptotics:3ndOrder:solutionQ3}
\dd{Q_3}{X} = K^2 Q_2 X \qquad 0 < X < 1.
\end{equation}
Inserting~\eqref{A:Asymptotics:3ndOrder:solutionQ3} into~\eqref{A:Asymptotics:3ndOrder:kinematic}, and using~\eqref{A:Asymptotics:2ndOrder:solQ2} gives
\begin{equation}\label{A:Asymptotics:3ndOrder:solutionsigma3}
\sigma_3  = -\frac{K^2}{6V^2}\left(2K^2 - 1\right).
\end{equation}
Undoing the various variable changes introduced in \S\ref{S:Asymptotics}, this gives the result in~\eqref{E:Asymptotics:SigmaResult}.

\begin{figure}
    \centering
    \includegraphics[width = 0.5\textwidth]{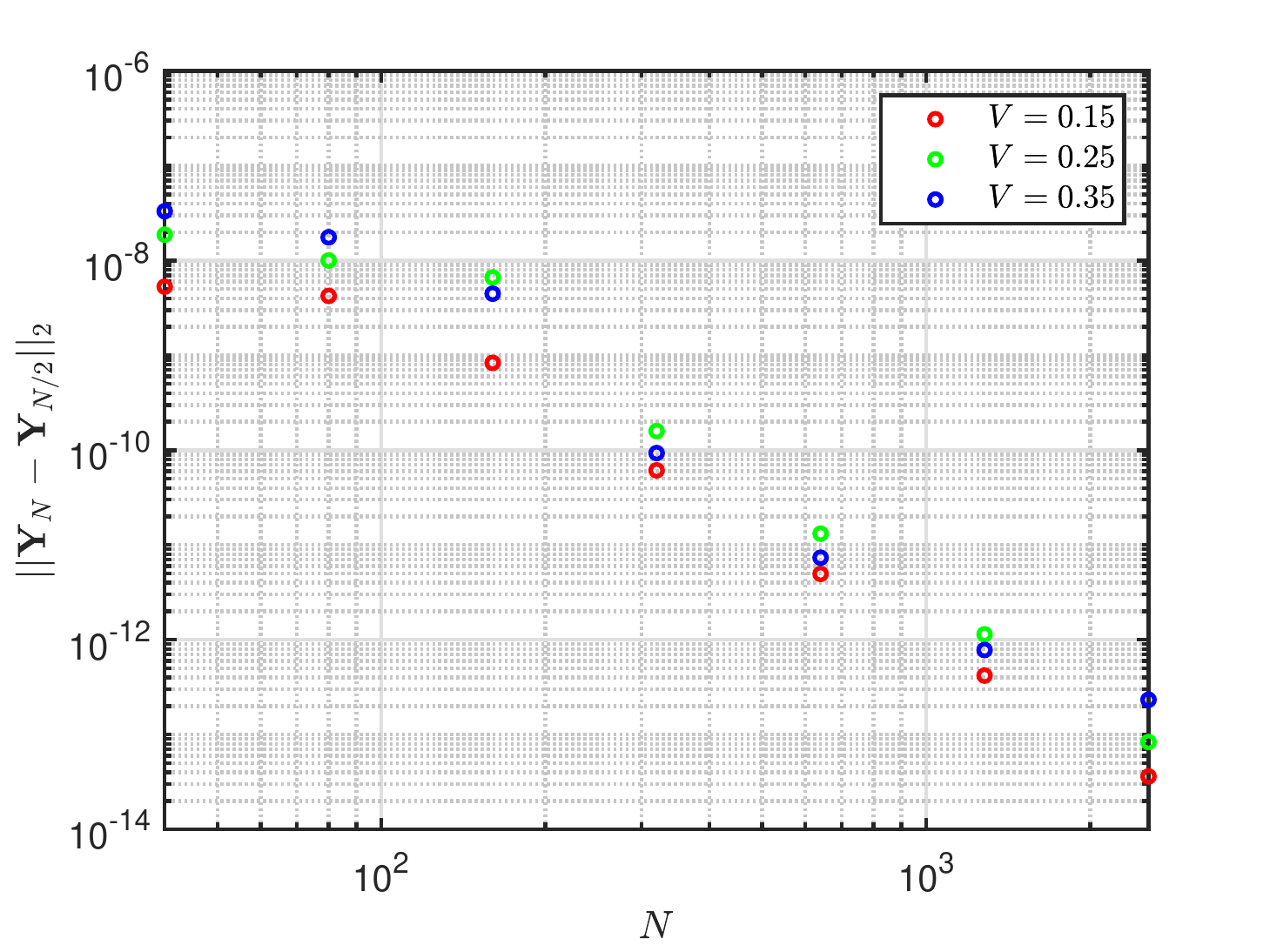}
    \caption{Difference between successive numerically obtained solutions of the BVP~\eqref{A:E:BVP} as a function of $N$, the number of grid points used in the numerical mesh. We show data for three different values of $V$, as indicated in the legend. All results presented here use $a = 0.01$ and $\bendability = 5$.}
    \label{fig:figA1_convergence}
\end{figure}

\blue{
\section{Numerical Scheme}\label{Appendix:Numerics}
In this section, we describe the numerical scheme used to solve the BVP~\eqref{E:LSA:ODEwet}--\eqref{E:LSA:kinematic}, which describes the linearized response of a periodic perturbation to an equilibrium configuration of the system.

To begin, we first find the appropriate equilibrium configuration using the procedure described in section~\ref{S:Equilibria}. As part of this procedure, the equilibrium meniscus position $x_0$ and meniscus channel width $h_0$ are also determined. We then express the system~\eqref{E:LSA:ODEwet}--\eqref{E:LSA:kinematic} as a multipoint BVP of the form
\begin{equation}\label{A:E:BVP}
\dd{\mathbf{Y}}{x} = \mathbf{f}(x; \sigma)
\end{equation}
where $\mathbf{Y} = (H, H',\dots, H^{\mathrm{IV}})^\intercal$ is a column vector containing $H$ and its first five derivatives, and $\sigma$ is the growth rate of the perturbation, which must be determined as part of the solution. This problem is referred to as `multipoint' because equation~\eqref{A:E:BVP} holds for all $0 < x < 1$, with the form of the right hand $\mathbf{f}$ varying depending on whether the wet ($0 < x < x_0$) or dry ($x_0 < x < 1$) region is appropriate at the particular value of $x$, and boundary conditions are applied at the interior point $x = x_0$, as well as at the domain boundaries, $x = 0$ and $x = 1$.

The BVP~\eqref{A:E:BVP} is solved numerically using the \texttt{BVP4C} routine implemented in \textsc{MATLAB}~\citep{Kierzenka2001BVP}. As part of this procedure, it is necessary to specify a numerical grid, an initial guess to the solution $\mathbf{Y}$ on that grid, and an initial guess at the growth rate $\sigma$. In all results shown in this paper, we chose a uniform grid containing a total of $N$ uniformly spaced grid points; in generating the data used in the figures in the main text, we take $N = 100$ and found that results are insensitive to further grid refinement. We take $\sigma = 0$ as an initial guess for $\sigma$ and $\mathbf{Y} = (1,1,1,1,1,1)^\intercal$ as an initial guess for $Y$. We found that solutions are insensitive to these initial guesses. The BVP~\eqref{A:E:BVP} is solved with a constant relative tolerance of $10^{-5}$ and an absolute tolerance $10^{-6}\times\bendability V^5 /a$, which scales with the size of the expected solutions [recall that $\bendability V^5 /a$ is the size of expected solution of the BVP~\eqref{A:E:BVP}, see section~\eqref{S:LSA}].

We verify that solutions obtained using the procedure described above converge as $N \to \infty$. Since the BVP~\eqref{A:E:BVP} does not have an analytic solution, to do so we consider the difference between successive approximations as a metric for convergence; this quantity decays in the limit $N \to \infty$ (figure~\ref{fig:figA1_convergence}), confirming that the procedure described above converges in this limit. }

\bibliographystyle{jfm}
\bibliography{mybib.bib}

\begin{thebibliography}{29}
\expandafter\ifx\csname natexlab\endcsname\relax\def\natexlab#1{#1}\fi
\def\au#1{#1} \def\ed#1{#1} \def\yr#1{#1}\def\at#1{#1}\def\jt#1{\textit{#1}}
  \def\bt#1{#1}\def\bvol#1{\textbf{#1}} \def\vol#1{#1} \def\pg#1{#1}
  \def\publ#1{#1}\def\arxiv#1{#1}\def\org#1{#1}\def\st#1{\textit{#1}}

\bibitem[Al-Housseiny \& Stone(2013)]{AlHousseiny2013PhysFlu}
{\sc \au{Al-Housseiny, T~T} \& \au{Stone, H~A}} \yr{2013}  \at{{Controlling
  viscous fingering in tapered Hele-Shaw cells}}.  \jt{Phys. Fluids}
  \bvol{25}~(9),  \pg{092102}.

\bibitem[Al-Housseiny {\em et~al.\/}(2012)Al-Housseiny, Tsai \&
  Stone]{AlHousseiny2012NaturePhysics}
{\sc \au{Al-Housseiny, T~T}, \au{Tsai, P~A} \& \au{Stone, H~A}} \yr{2012}
  \at{{Control of interfacial instabilities using flow geometry}}.  \jt{Nat.
  Phys.}  \bvol{8}~(10),  \pg{747--750}.

\bibitem[Aristoff {\em et~al.\/}(2011)Aristoff, Duprat \&
  Stone]{Aristoff2011IntJNonlinMech}
{\sc \au{Aristoff, J~M}, \au{Duprat, C} \& \au{Stone, H~A}} \yr{2011}
  \at{{Elastocapillary imbibition}}.  \jt{Int. J. Non-Linear Mech.}
  \bvol{46}~(4),  \pg{648--656}.

\bibitem[Bradley(2020)]{BradleyPhDthesis}
{\sc \au{Bradley, A~T}} \yr{2020}  \at{Droplet transport by bendotaxis}. PhD
  thesis, University of Oxford.

\bibitem[Bradley(2022)]{BendocapillaryRepo}
{\sc \au{Bradley, A~T}} \yr{2022} Code to run simulations and produce figures.
  \url{https://github.com/alextbradley/Bendocapillary}, accessed: 2022-01-04.

\bibitem[Bradley {\em et~al.\/}(2019)Bradley, Box, Hewitt \&
  Vella]{Bradley2019PRL}
{\sc \au{Bradley, A~T}, \au{Box, F}, \au{Hewitt, I~J} \& \au{Vella, D}}
  \yr{2019}  \at{{Wettability-Independent Droplet Transport by Bendotaxis}}.
  \jt{Phys. Rev. Lett.}  \bvol{122}~(7),  \pg{074503}.

\bibitem[Bradley {\em et~al.\/}(2021)Bradley, Hewitt \& Vella]{Bradley2021PRF}
{\sc \au{Bradley, A~T}, \au{Hewitt, I~J} \& \au{Vella, D}} \yr{2021}
  \at{Droplet trapping in bendotaxis caused by contact angle hysteresis}.
  \jt{Phys. Rev. Fluids}  \bvol{6},  \pg{114003}.

\bibitem[Duprat {\em et~al.\/}(2011)Duprat, Aristoff \& Stone]{Duprat2011JFM}
{\sc \au{Duprat, C}, \au{Aristoff, J~M} \& \au{Stone, H~A}} \yr{2011}
  \at{{Dynamics of elastocapillary rise}}.  \jt{J. Fluid Mech.}  \bvol{679},
  \pg{641--654}.

\bibitem[Eggers \& Villermaux(2008)]{Eggers2008RepProgPhys}
{\sc \au{Eggers, J} \& \au{Villermaux, E}} \yr{2008}  \at{Physics of liquid
  jets}.  \jt{Rep. Prog. Phys.}  \bvol{71}~(3),  \pg{036601}.

\bibitem[Flitton \& King(2004)]{Flitton2004EJApplMech}
{\sc \au{Flitton, J.~C.} \& \au{King, J.~R.}} \yr{2004}  \at{Moving-boundary
  and fixed-domain problems for a sixth-order thin-film equation}.  \jt{Eur. J.
  Appl. Math.}  \bvol{15}~(6),  \pg{713–754}.

\bibitem[F{\"o}ppl(1921)]{Foppl1921}
{\sc \au{F{\"o}ppl, A}} \yr{1921} {\em Vorlesungen {\"u}ber technische
  Mechanik\/}, ,  \vol{vol.~6}.  \publ{B. G. Teubner}.

\bibitem[Ha {\em et~al.\/}(2021)Ha, Kim, Siu \& Tawfick]{Ha2021SoftMatter}
{\sc \au{Ha, J}, \au{Kim, Y~S}, \au{Siu, R} \& \au{Tawfick, S}} \yr{2021}
  \at{Dynamic pattern selection in polymorphic elastocapillarity}.  \jt{Soft
  Matter} .

\bibitem[Hadjittofis {\em et~al.\/}(2016)Hadjittofis, Lister, Singh \&
  Vella]{Hadjittofis2016JFM}
{\sc \au{Hadjittofis, A}, \au{Lister, J~R}, \au{Singh, K} \& \au{Vella, D}}
  \yr{2016}  \at{{Evaporation effects in elastocapillary aggregation}}.  \jt{J.
  Fluid Mech.}  \bvol{792},  \pg{168--185}.

\bibitem[von K{\'a}rm{\'a}n(1910)]{Karman1907}
{\sc \au{von K{\'a}rm{\'a}n, T}} \yr{1910}  \at{Festigkeitsprobleme im
  maschinenbau}.  \jt{Encyk D Math Wiss IV}  \pg{pp. 311--385}.

\bibitem[Keiser {\em et~al.\/}(2016)Keiser, Herbaut, Bico \&
  Reyssat]{Keiser2016JFM}
{\sc \au{Keiser, L}, \au{Herbaut, R}, \au{Bico, J} \& \au{Reyssat, E}}
  \yr{2016}  \at{{Washing wedges: capillary instability in a gradient of
  confinement}}.  \jt{J. Fluid Mech.}  \bvol{790},  \pg{619--633}.

\bibitem[Kierzenka \& Shampine(2001)]{Kierzenka2001BVP}
{\sc \au{Kierzenka, J} \& \au{Shampine, L~F}} \yr{2001}  \at{{A BVP solver
  based on residual control and the MATLAB PSE}}.  \jt{ACM Trans. Math. Softw.}
   \bvol{27}~(3),  \pg{299--316}.

\bibitem[Langer(1980)]{Langer1980RevModPhys}
{\sc \au{Langer, J~S}} \yr{1980}  \at{Instabilities and pattern formation in
  crystal growth}.  \jt{Rev. Mod. Phys.}  \bvol{52}~(1),  \pg{1}.

\bibitem[Leal(2007)]{Leal2007}
{\sc \au{Leal, L~Gary}} \yr{2007} {\em Advanced transport phenomena: fluid
  mechanics and convective transport processes\/}, ,  \vol{vol.~7}.
  \publ{Cambridge University Press}.

\bibitem[Ledesma-Aguilar {\em et~al.\/}(2017)Ledesma-Aguilar, Laghezza, Yeomans
  \& Vella]{LedesmaAguilar2017SoftMatter}
{\sc \au{Ledesma-Aguilar, R}, \au{Laghezza, G}, \au{Yeomans, J~M} \& \au{Vella,
  D}} \yr{2017}  \at{{Using evaporation to control capillary instabilities in
  micro-systems}}.  \jt{Soft Matter}  \bvol{13}~(47),  \pg{8947--8956}.

\bibitem[Plateau(1873)]{Plateau1873}
{\sc \au{Plateau, J}} \yr{1873} {\em {Statique exp{\'e}rimentale et
  th{\'e}or{\'e}tique des liquides soumis aux seules forces
  mol{\'e}culaires}\/}.  \publ{Paris: Gautier-Villars}.

\bibitem[Rayleigh(1879)]{Rayleigh1879PRSL}
{\sc \au{Rayleigh, L}} \yr{1879}  \at{{On the capillary phenomena of jets}}.
  \jt{Proc. R. Soc. London}  \bvol{29}~(71).

\bibitem[Rayleigh(1892)]{Rayleigh1892PhilosMag}
{\sc \au{Rayleigh, L}} \yr{1892}  \at{{On the instability of a cylinder of
  viscous liquid under capillary force}}.  \jt{Philos. Mag.}  \bvol{34}~(207),
  \pg{145--154}.

\bibitem[Reddy(2006)]{Reddy2006}
{\sc \au{Reddy, J~N}} \yr{2006} {\em Theory and analysis of elastic plates and
  shells\/}.  \publ{CRC press}.

\bibitem[Reyssat(2014)]{Reyssat2014JFM}
{\sc \au{Reyssat, E}} \yr{2014}  \at{{Drops and bubbles in wedges}}.  \jt{J.
  Fluid Mech.}  \bvol{748},  \pg{641--662}.

\bibitem[Saffman \& Taylor(1958)]{Saffman1958PRSL}
{\sc \au{Saffman, P~G} \& \au{Taylor, G~I}} \yr{1958}  \at{{The penetration of
  a fluid into a porous medium or Hele-Shaw cell containing a more viscous
  liquid}}.  \jt{Proceedings of the Royal Society of London}
  \bvol{245}~(1242),  \pg{312--329}.

\bibitem[Seemann {\em et~al.\/}(2011)Seemann, Brinkmann, Herminghaus, Khare,
  Law, McBride, Kostourou, Gurevich, Bommer, Herrmann \&
  Michler]{Seemann2011JPhysCondMat}
{\sc \au{Seemann, R}, \au{Brinkmann, M}, \au{Herminghaus, S}, \au{Khare, K},
  \au{Law, B~M}, \au{McBride, S}, \au{Kostourou, K}, \au{Gurevich, E},
  \au{Bommer, S}, \au{Herrmann, C} \& \au{Michler, D}} \yr{2011}  \at{{Wetting
  morphologies and their transitions in grooved substrates}}.  \jt{J. Phys.
  Condens. Matter}  \bvol{23}~(18),  \pg{184108--17}.

\bibitem[Taroni \& Vella(2012)]{Taroni2012JFM}
{\sc \au{Taroni, M} \& \au{Vella, D}} \yr{2012}  \at{{Multiple equilibria in a
  simple elastocapillary system}}.  \jt{J. Fluid Mech.}  \bvol{712},
  \pg{273--294}.

\bibitem[Timoshenko \& Woinowsky-Krieger(1959)]{Timoshenko1959}
{\sc \au{Timoshenko, S.} \& \au{Woinowsky-Krieger, S.}} \yr{1959} {\em Theory
  of Plates and Shells\/}.  \publ{McGraw-Hill}.

\bibitem[Zheng {\em et~al.\/}(2015)Zheng, Kim \& Stone]{Zheng2015PRL}
{\sc \au{Zheng, Z}, \au{Kim, H} \& \au{Stone, H~A}} \yr{2015}  \at{Controlling
  viscous fingering using time-dependent strategies}.  \jt{Phys. Rev. Lett.}
  \bvol{115}~(17),  \pg{174501}.

\end{thebibliography}

\end{document}